\documentclass[twocolumn]{aastex63}

\usepackage{bm}

\newcommand{\fice}{f_\mathrm{ice}}
\newcommand{\tauabs}{\tau_\mathrm{abs}}

\newcommand{\amax}{a_\mathrm{max}}
\newcommand{\amin}{a_\mathrm{min}}
\newcommand{\unitden}{\rm{g}~\rm{cm}^{-3}}
\newcommand{\amean}{\langle{a}\rangle}
\shorttitle{Scattering-polarization feature of water-ice grains}
\shortauthors{Tazaki et al.}

\begin{document}

\title{Scattering polarization of 3-$\mu$m water-ice feature by large icy grains}

\correspondingauthor{Ryo Tazaki}
\email{r.tazaki@uva.nl}

\author[0000-0003-1451-6836]{Ryo Tazaki}
\affil{Anton Pannekoek Institute for Astronomy, University of Amsterdam, Science Park 904, 1098XH Amsterdam, The Netherlands}
\affil{Division of Liberal Arts, Kogakuin University, 1-24-2 Nishi-Shinjuku, Shinjuku-ku, Tokyo 163-8677, Japan}
\affil{Astronomical Institute, Graduate School of Science,
Tohoku University, 6-3 Aramaki, Aoba-ku, Sendai 980-8578, Japan}

\author[0000-0001-9843-2909]{Koji Murakawa}
\affil{Institute of Education, Osaka Sangyo University, 3-1-1 Nakagaito, Daito, Osaka 574-8530, Japan}

\author{Takayuki Muto}
\affil{Division of Liberal Arts, Kogakuin University, 1-24-2 Nishi-Shinjuku, Shinjuku-ku, Tokyo 163-8677, Japan}

\author[0000-0002-6172-9124]{Mitsuhiko Honda}
\affil{Faculty of Biosphere-Geosphere Science, Okayama University of Science, 1-1 Ridai-chou, Okayama 700-0005, Japan}

\author[0000-0002-7779-8677]{Akio K. Inoue}
\affil{Department of Physics, School of Advanced Science and Engineering, Faculty of Science and Engineering, Waseda University, 3-4-1, Okubo, Shinjuku, Tokyo 169-8555, Japan}
\affil{Waseda Research Institute for Science and Engineering, Faculty of Science and Engineering, Waseda University, 3-4-1, Okubo, Shinjuku, Tokyo 169-8555, Japan}




\begin{abstract}
Water ice has a strong spectral feature at a wavelength of approximately $3~\mu$m, which plays a vital role in our understanding of the icy universe. In this study, we investigate the scattering polarization of this water-ice feature. The linear polarization degree of light scattered by $\mu$m-sized icy grains is known to be enhanced at the ice band; however, the dependence of this polarization enhancement on various grain properties is unclear. 
We find that the enhanced polarization at the ice band is sensitive to the presence of $\mu$m-sized grains as well as their ice abundance. 
We demonstrate that this enhancement is caused by the high absorbency of the water-ice feature, which attenuates internal scattering and renders the surface reflection dominant over internal scattering.
Additionally, we compare our models with polarimetric observations of the low-mass protostar L1551 IRS 5. Our results show that scattering by a maximum grain radius of a few microns with a low water-ice abundance is consistent with observations. Thus, scattering polarization of the water-ice feature is a useful tool for characterizing ice properties in various astronomical environments. 
\end{abstract}



\section{Introduction}\label{sec:intro}
Water ice is the most abundant volatile in star- and planet-forming regions \citep{Pollack94}, and its importance in these regions is manifold. Water ice provides a site for efficient chemical reactions, leading to the formation of complex molecules \citep{Herbst09}. Its high adhesion may also facilitate planet formation \citep{Gundlach11, Gundlach15, wada09, Wada13}. Therefore, ice evolution intimately links physical and chemical evolution in these regions.

Water ice has a strong spectral feature at a wavelength of approximately $3~\mu$m, which is attributed to the O--H vibration of water molecules in ice. The 3-$\mu$m feature of water ice has been ubiquitously detected in star- and planet-forming regions \citep[e.g.,][]{Boogert15} as either an absorption feature \citep{Whittet01, Pont05, Terada07, Terada12a, Terada12b, Terada17} or a scattering feature \citep{Pendleton90, Honda09, Honda16}. However, the grain-size distribution and ice abundance are often degenerate, complicating the interpretation. Therefore, additional observations of the 3-$\mu$m feature are desirable.

Scattering polarization of the 3-$\mu$m feature may shed light on both the grain-size distribution and ice abundance. \citet{Pendleton90} found that light scattered by $\mu$m-sized icy grains shows an enhanced degree of linear polarization at the ice band. Because this polarization enhancement does not occur for sub-micron-sized grains, e.g., interstellar grains \citep{MRN77}, this enhancement may indicate the presence of $\mu$m-sized icy grains.

\citet{Kobayashi99} detected polarization enhancement at the ice band in the Class I protostar L1551 IRS 5, hinting at the presence of $\mu$m-sized icy grains in the protostar envelope. 
Although it has not yet been confirmed, the polarization enhancement might be seen in protoplanetary disks as disk scattered light of the ice feature has been detected \citep{Honda09, Honda16}. In addition, debris disks are also anticipated to exhibit the polarization enhancement \citep{Kim19}. Therefore, scattering polarization of the ice feature might be potentially observable for various star- and planet-forming regions and useful for constraining ice properties. 

Despite its potential importance, the dependence of this polarization enhancement on various grain properties is still unexplored. In fact, the detailed ice grain properties needed to explain the observations by \citet{Kobayashi99} remains unclear. Thus, in this study, we augment the work of \citet{Pendleton90} by systematically studying scattering polarization at the ice band for various grain-size distributions, scattering angles, and ice abundance and show how the scattering polarization looks like across various environments in star- and planet-forming regions. In addition, we apply our model calculation to polarimetric observations of the envelope of the low-mass protostar L1551 IRS 5 and investigate ice properties necessary to explain the observations.

This paper is organized as follows. 
In Section \ref{sec:scatpol}, we summarize the dust model adopted in this study and explore the dependence of scattering polarization on grain-size distribution. We investigate the physical origin of this polarization enhancement in Section \ref{sec:origin}. 
In Section \ref{sec:obs}, we compare our models with observations of the low-mass protostar L1551 IRS 5. In Section \ref{sec:disc}, we discuss potential applications of the scattering polarization of the ice band for various astronomical objects. We conclude in Section \ref{sec:summary}.

\section{Scattering-polarization feature of water ice} \label{sec:scatpol}
\begin{figure*}[htbp]
\begin{center}
\includegraphics[height=6.0cm,keepaspectratio]{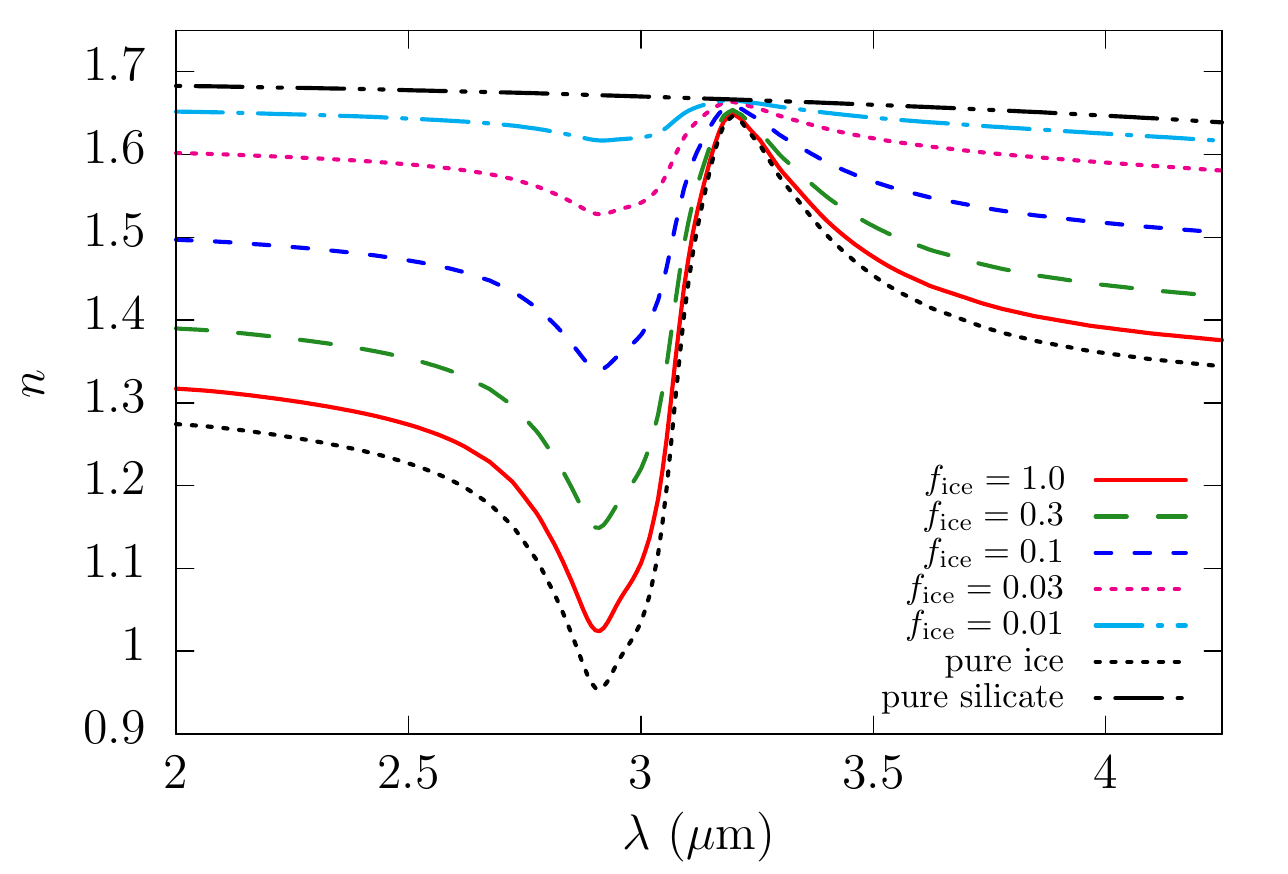}
\includegraphics[height=6.0cm,keepaspectratio]{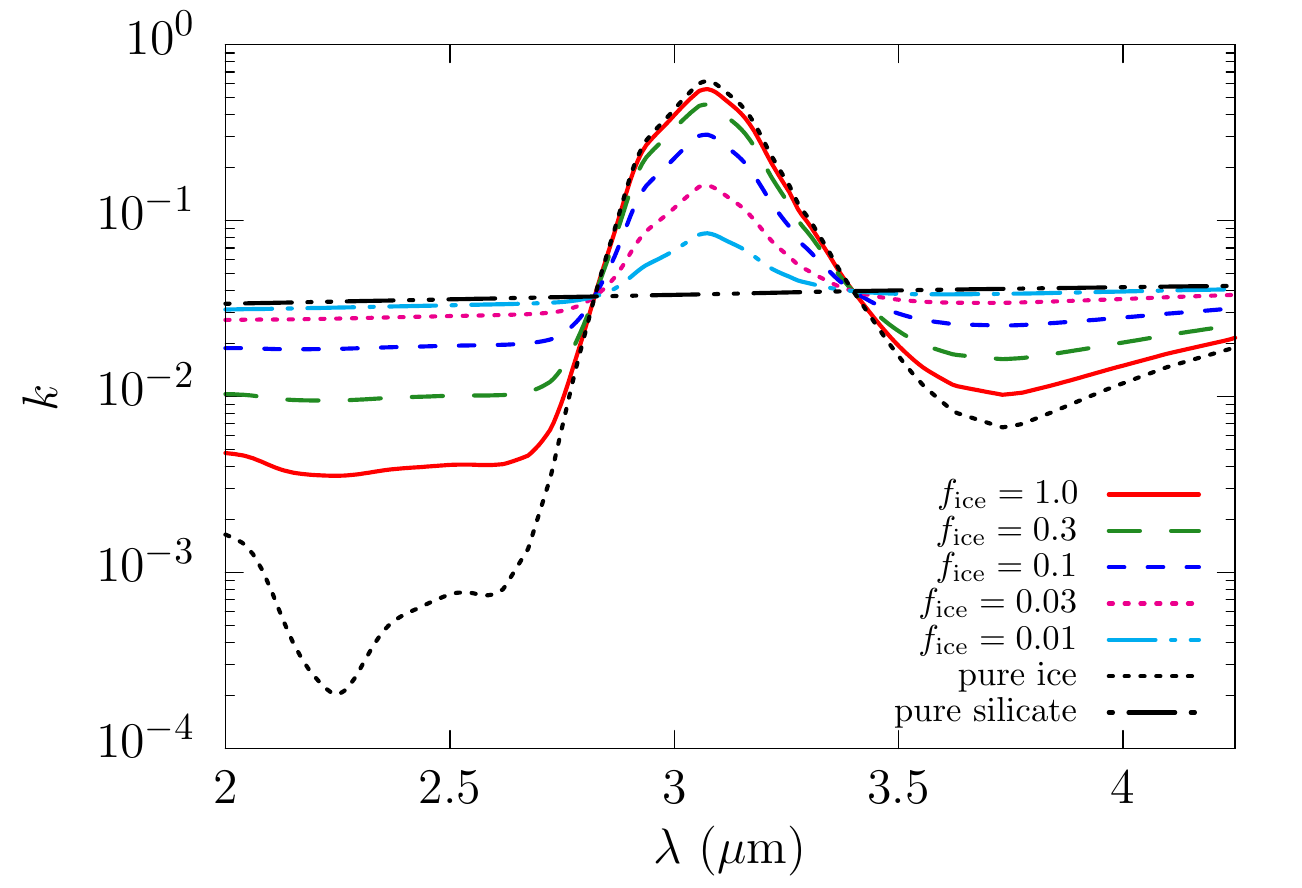}
\caption{Refractive indices, $m=n+ik$, of a silicate-ice mixture at the 3-$\mu$m water-ice feature. The left and right panels show the real and imaginary part of the complex refractive index, respectively. In each panel, the refractive index is shown for various values of $\fice$ as well as that of pure silicate and pure ice.}
\label{fig:opcont}
\end{center}
\end{figure*}

\subsection{Dust models}
We assume that each grain consists of silicate and water ice. 
The mass abundances of silicate (olivine) and water ice are assumed to be $\zeta_\mathrm{sil}=\zeta_\mathrm{sil}^{\mathrm{P94}}$ and $\zeta_\mathrm{ice}=\fice \zeta_\mathrm{ice}^{\mathrm{P94}}$, where $\fice$ is a free parameter related to the water-ice abundance. $\zeta_\mathrm{sil}^{\mathrm{P94}}=2.64\times10^{-3}$ and $\zeta_\mathrm{ice}^{\mathrm{P94}}=5.55\times10^{-3}$ are the silicate and water-ice abundances proposed by \citet{Pollack94}. 
The material densities of silicate and water ice are also taken from \citet{Pollack94}, as $\rho_\mathrm{sil}=3.5~\unitden$ and $0.92~\unitden$, respectively. By using their mass abundances ($\zeta_\mathrm{sil}$ and $\zeta_\mathrm{ice}$) and material densities, we derive the volume fraction of each component for various values of $\fice$, as summarized in Table \ref{tab:fvol}. 
\begin{table}[tbp]
  \caption{Mass Fraction (Volume Fraction) of Water Ice and Silicate in Icy Grains}
  \label{tab:fvol}
  \centering
  \begin{tabular}{lcc}
    \hline
    Model & Water Ice & Silicate \\
    \hline \hline
    $f_\mathrm{ice}=1$ & 68\% (89\%) & 32\% (11\%)  \\
    $f_\mathrm{ice}=0.3$ & 39\% (71\%) & 61\% (29\%) \\    
    $f_\mathrm{ice}=0.1$ & 17\% (44\%) & 83\% (56\%)  \\
    $f_\mathrm{ice}=0.03$ & 6\% (19\%) & 94\% (81\%)  \\
    $f_\mathrm{ice}=0.01$ & 2\% (7\%)& 98\% (93\%) \\
    \hline
  \end{tabular}
\end{table}

We calculate the scattering matrix elements for the grains using the Mie theory with an effective medium approximation. The refractive indices of silicate and water ice are taken from \citet{Draine03b} and \citet{Warren08}, respectively.
The refractive indices of the components are mixed according to Bruggeman's mixing rule \citep{Bruggeman35}, as shown in Figure \ref{fig:opcont}. Once the refractive index is obtained, we use the Mie theory to find the scattering matrix elements \citep{Bohren83}. A core-mantle grain structure and compositionally segregated grains are other potential forms of a silicate-ice mixture. We have computed the values for these forms and found that these approaches tend to produce results that are qualitatively similar to those obtained by the effective medium approach. 
Thus, we adopt the effective medium approach in this paper.

Scattering matrix elements of single-sized grains are averaged over the grain-size distribution. 
The distribution-averaged scattering matrix elements, $\langle S_{ij} \rangle$, are given by
\begin{equation}
\langle S_{ij}(\theta)\rangle=\frac{\int S_{ij}(a,\theta)n(a)da}{\int n(a)da}, \label{eq:dist}
\end{equation}
where $a$ is the grain radius, $\theta$ is the scattering angle, $S_{ij}$ is the scattering matrix element of single-sized grains, and $n(a)$ is the size distribution. Hereafter, we omit the bracket symbol in the averaged quantity, unless it is ambiguous.

The grain-size distribution is assumed to obey a single power-law function:
\begin{eqnarray}
n(a)da\propto\left\{ \begin{array}{ll}
a^{-q}da & (\amin\le a \le \amax), \\
0 & (\mathrm{otherwise}), \\
\end{array} \right.
\end{eqnarray}
where $n(a)da$ is the number density of grains with radii between $a$ and $a+da$ and $\amin$ and $\amax$ are the minimum and maximum grain radii, respectively. 

Our primary focus is scattered light observations of icy particles in star- and planet-forming regions. Hence, we vary $\amax$ from $0.1~\mu$m to 1 mm and $\amin$ from $0.005~\mu$m to $100~\mu$m to cover a wide range of parameter space expected in molecular clouds \citep[e.g.,][]{Pagani10, Steinacker10}, protoplanetary disks \citep[e.g.,][]{Mulders13, Tazaki19}, and debris disks \citep[e.g.,][]{Mittal15, Hughes18}. 
Although we treat $q=3.5$ as a fiducial case \citep[e.g.,][]{Dohnanyi69, Tanaka96}, we also investigate how different values of $q$ affect scattering polarization. We consider $q=2.5$ and $3.0$ because the scattering property of a distribution of icy particles changes at $q=3.0$. 

\subsection{Brief summary of the scattering-polarization feature of water ice}
Before presenting a detailed discussion, we summarize some characteristics of the scattering polarization of the 3-$\mu$m water-ice feature, including some updates obtained in this study.

\begin{figure}[t]
\begin{center}
\includegraphics[width=1.0\linewidth,keepaspectratio]{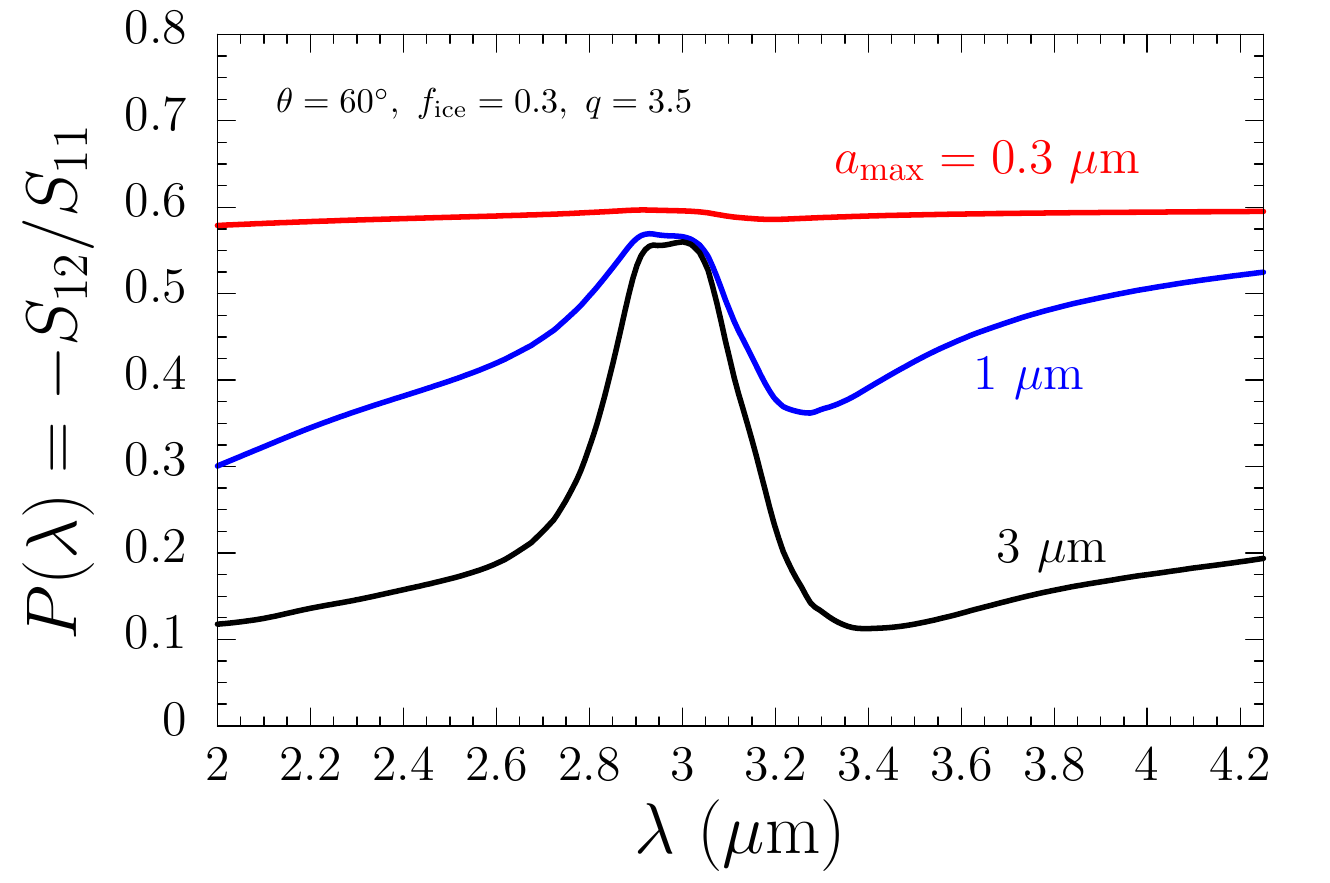}
\caption{Scattering-polarization features of water ice with $\fice=0.3$ at $\theta=60^\circ$. The red, blue, and black lines present results for $\amax=0.3~\mu$m, $1~\mu$m, and $3~\mu$m, respectively. The minimum grain radius and power-law index of the size distribution are $\amin=0.005~\mu$m and $q=3.5$, respectively.}
\label{fig:polwave}
\end{center}
\end{figure}

Figure \ref{fig:polwave} shows the linear polarization degree of scattered light as a function of wavelength for the 3-$\mu$m water-ice feature. This figure clearly demonstrates that the wavelength dependence of the polarization degree is sensitive to $\amax$. 
When $\amax=0.3~\mu$m, the degree of polarization is almost independent of wavelength.
When $\amax=1~\mu$m, the polarization degree varies with wavelength and exhibits an excess at $\lambda\sim3~\mu$m. 
When $\amax=3~\mu$m, the polarization excess becomes more pronounced. Thus, larger grains tend to show a more prominent polarization excess. We will refer to this polarization excess as the {\it scattering-polarization feature} or simply the polarization feature.

\begin{figure}[t]
\begin{center}
\includegraphics[width=1.0\linewidth,keepaspectratio]{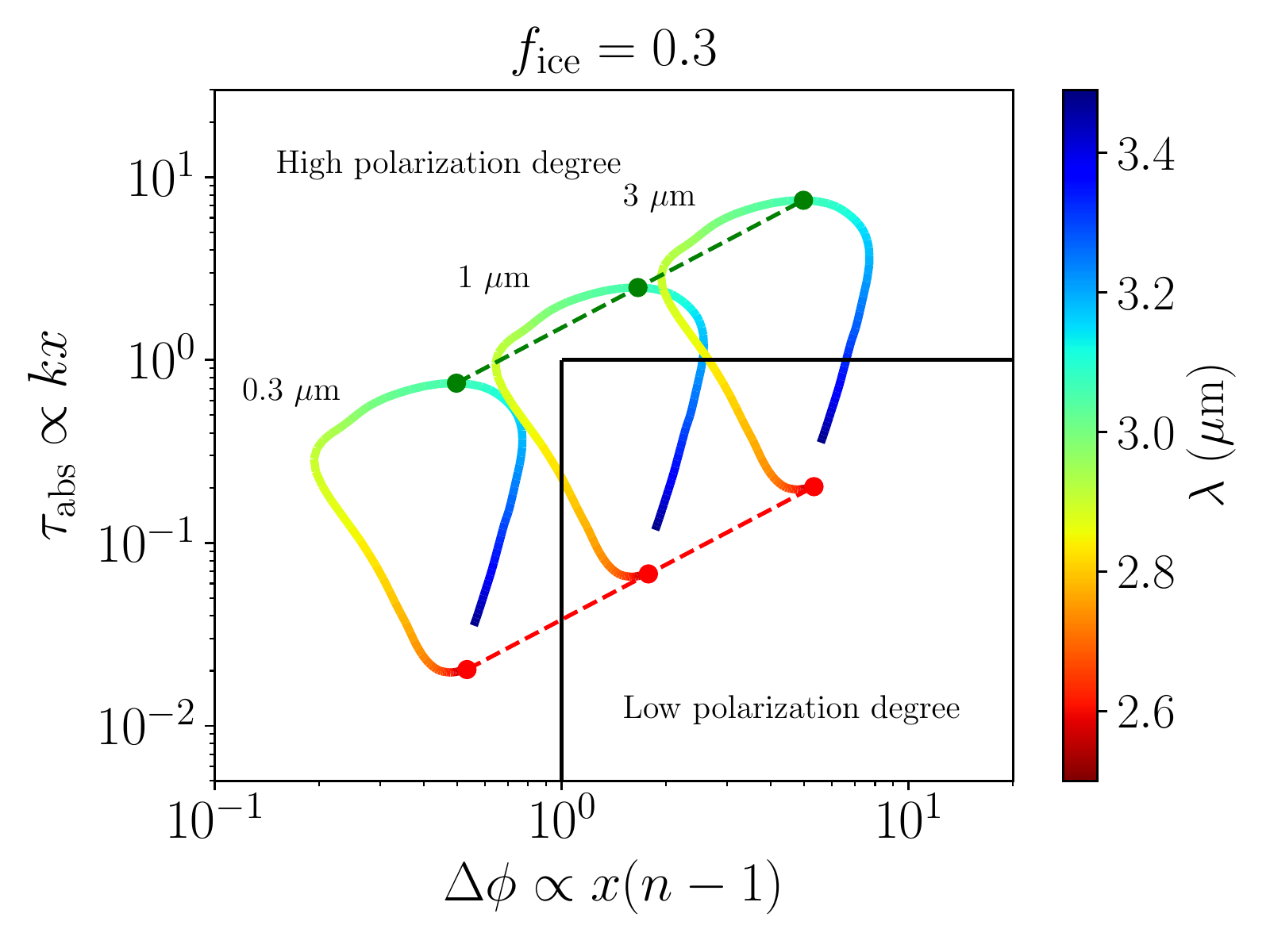}
\caption{The phase shift $\Delta\phi$ and absorption optical depth $\tauabs$ for an ice-silicate mixture ($\fice=0.3$). Each line presents the result obtained for a single grain radius ($a=0.3~\mu$m, 1 $\mu$m, and 3 $\mu$m), and along each line, the wavelength varies from $\lambda=2.5~\mu$m to $3.5~\mu$m, as shown in the color scale. The red and green dashed lines indicate constant wavelengths of $\lambda=2.5~\mu$m and $3.07~\mu$m, respectively. The lower-right region tends to produce a low degree of polarization due to multiple scattering, whereas the remaining regions tend to produce a high degree of polarization due to either single scattering or reflection.}
\label{fig:regime}
\end{center}
\end{figure}

Scattered light polarization can be characterized by two parameters: (i) the phase shift $\Delta\phi=2x(n-1)$, where $n$ is the real part of the refractive index, $x=2\pi{a}/\lambda$ is the size parameter, and $\lambda$ is the wavelength \citep{Bohren83}, and (ii) the absorption optical depth of a dust grain, $\tauabs=8kx/3$, where $k$ is the imaginary part of the refractive index \citep{Kataoka14}. The polarization can be classified into three regimes, according to the behavior of the incident light.
\begin{itemize}
    \item Single-scattering regime ($\Delta\phi\ll1$): Rayleigh or Rayleigh-Gans scattering occurs \citep{Bohren83}. 
  The polarization degree in this case is high, i.e., 100\% at a scattering angle of $90^\circ$. 
    \item Multiple-scattering regime ($\Delta\phi\gtrsim1$ and $\tauabs\lesssim1$): 
    The incident light is scattered multiple times inside the grain without experiencing significant absorption. The degree of linear polarization is generally (but not always) smaller than that of the single-scattering case.
    Negative polarization ($P=-S_{12}/S_{11}<0$) is also often observed in this regime.
    \item Reflection regime ($\tauabs\gg1$): 
    The light that penetrates the grain is absorbed. Because the size parameter is generally $\gtrsim 1$ in this regime (because $k\lesssim1$), the surface Fresnel reflection is dominant. Due to this Fresnel reflection, the polarization degree of surface-reflected light is high, i.e., 100\% at the Brewster angle.
\end{itemize}

The polarization feature can be understood by applying the $\Delta\phi-\tauabs$ plane. 
Figure \ref{fig:regime} shows how $\Delta\phi$ and $\tauabs$ vary in the wavelength range of the ice band. Rapid changes in the refractive index at the ice band create a loop-like trajectory in the $\Delta\phi-\tauabs$ plane, which is a characteristic of a Lorentz oscillator \citep{Bohren83}. The polarization degree drastically changes at the boundaries of the multiple-scattering regime. For $a=0.3~\mu$m, the trajectory does not cross the boundaries of the multiple-scattering regime; hence, the wavelength dependence of the polarization is weak. For $\mu$m-sized grains, the short- and long-wavelength ends of the ice band are located in the multiple-scattering regime, whereas the central wavelength is located in the reflection regime. As a result, the polarization degree is enhanced near the central wavelength of the ice band. 

\citet{Pendleton90} argued that the polarization feature is caused by a trajectory crossing the plane of $\Delta\phi=1$. Here, we argue that both $\Delta\phi$ and $\tauabs$ play an important role in forming the polarization feature. We clarify the role of the absorption optical depth for polarization in Section \ref{sec:origin}.

\subsection{Effect of size distribution on the polarization feature}
Here, we examine how the polarization feature depends on various parameters of the grain-size distribution: $\amax$, $\amin$, and $q$.

\subsubsection{Polarization property of single-sized grains} \label{sec:single}
\begin{figure}[t]
\begin{center}
\includegraphics[width=1.0\linewidth,keepaspectratio]{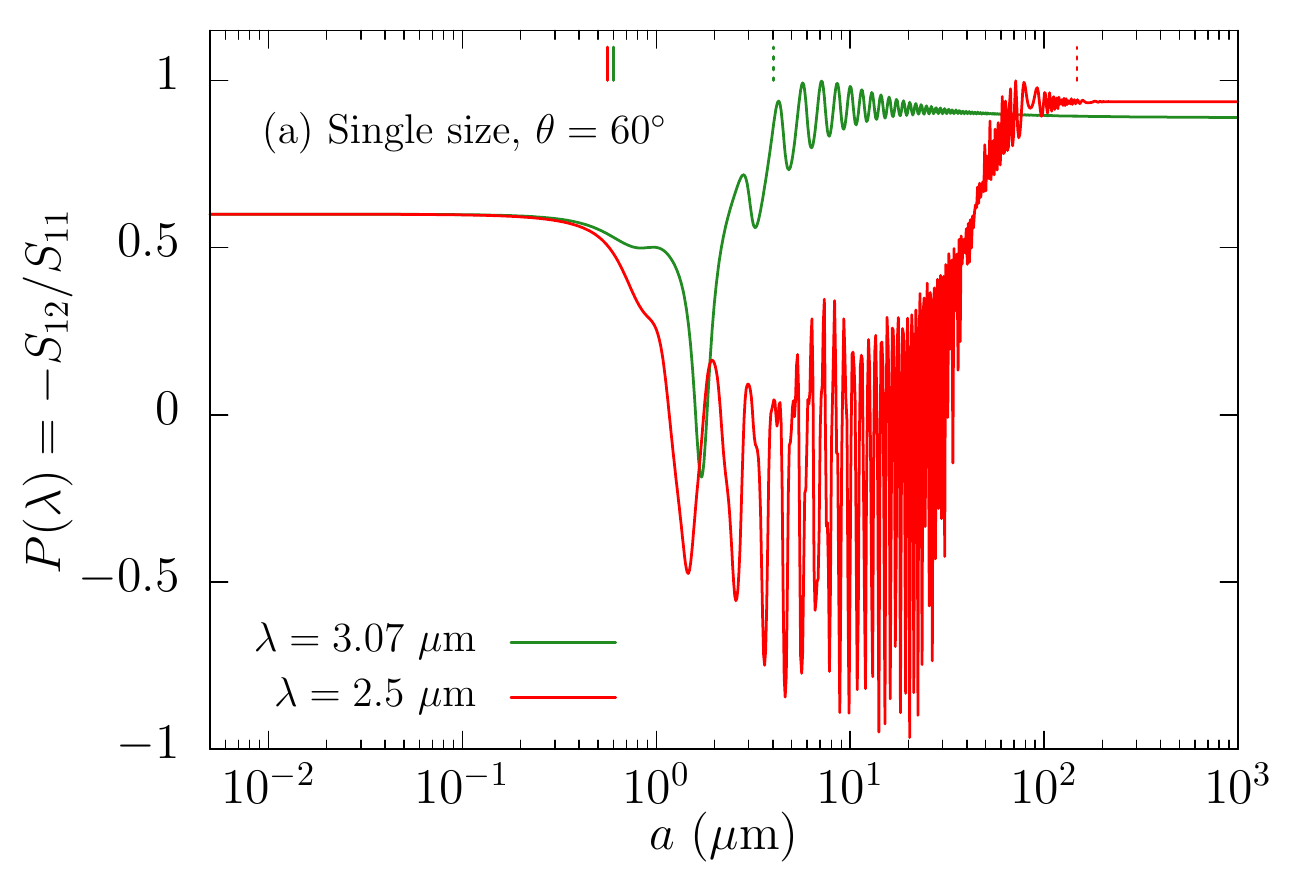}
\includegraphics[width=1.0\linewidth,keepaspectratio]{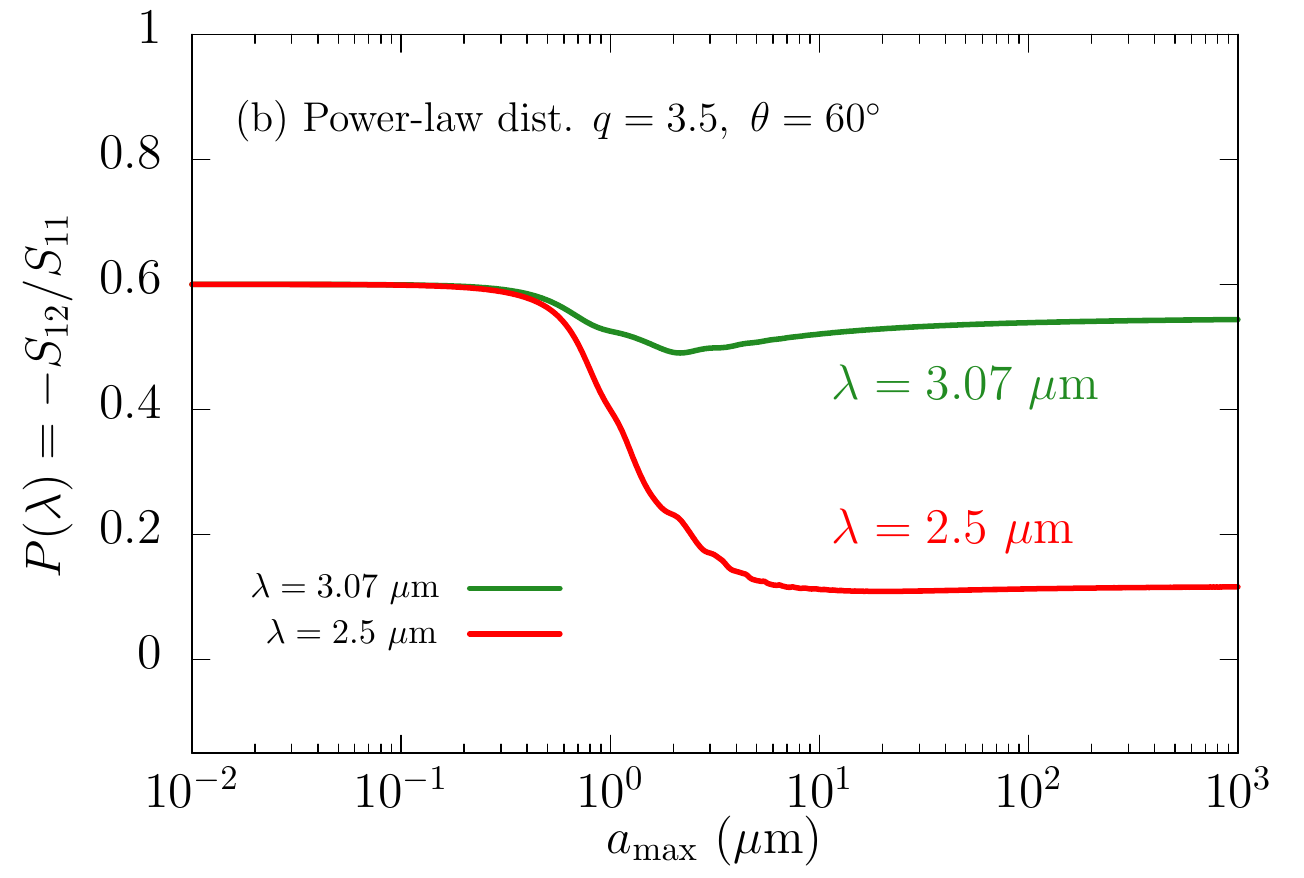}
\includegraphics[width=1.0\linewidth,keepaspectratio]{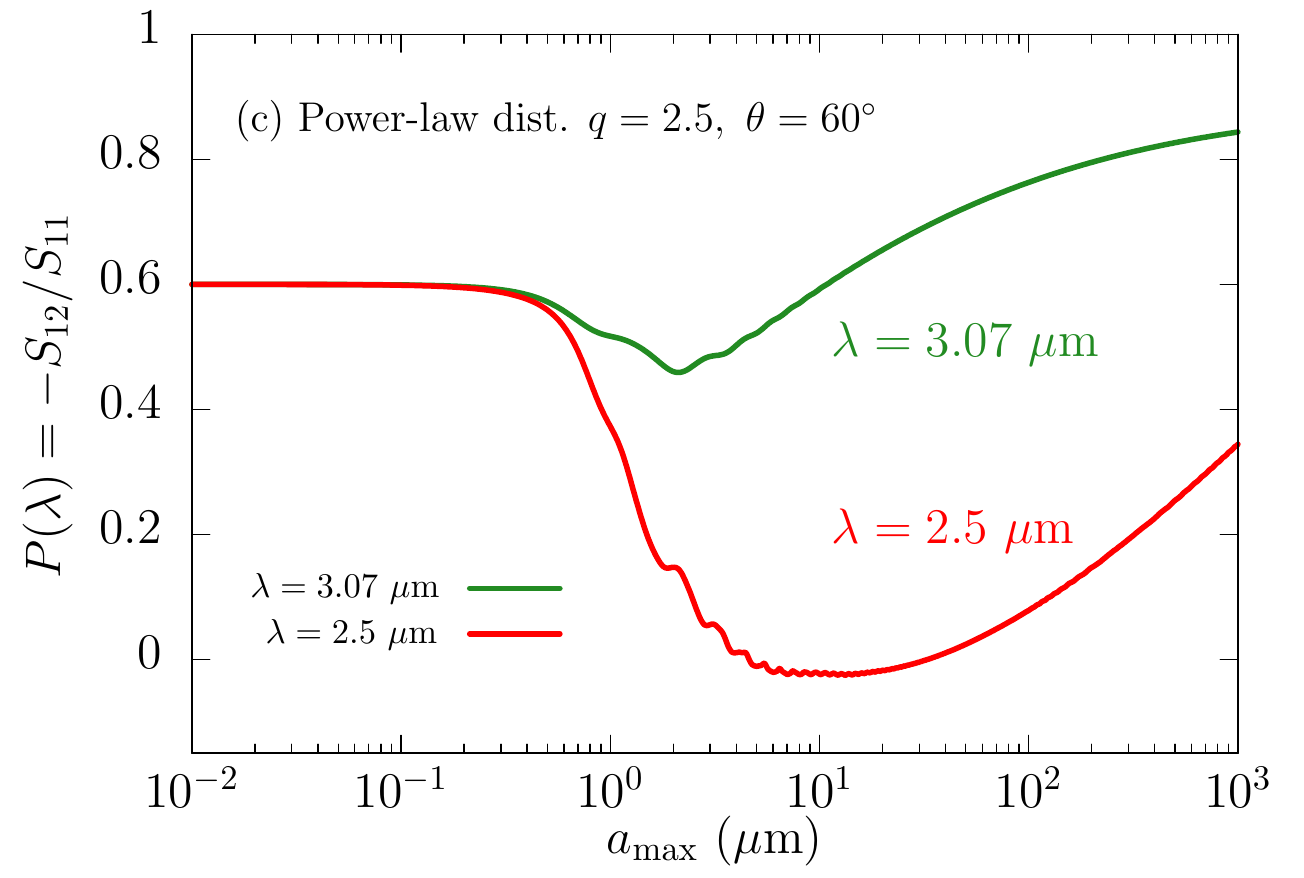}
\caption{ (a) Red and green lines show the degree of polarization at $\lambda=2.5~\mu$m and $3.07~\mu$m, respectively. The scattering angle is fixed at $\theta=60^\circ$. Vertical solid and dotted lines show the grain radius at $\Delta\phi=1$ and $\tauabs=10$, respectively. (b, c) Same as (a), but for the distribution-averaged polarization degree with $q=3.5$ (b) and $q=2.5$ (c).}
\label{fig:polsize}
\end{center}
\end{figure}

Figure \ref{fig:polsize} (top) shows the polarization degree of single-sized grains as a function of radius, which provides a useful starting point for our discussion.
We compare the degree of polarization at the center ($\lambda=3.07~\mu$m) and at the outside ($\lambda=2.5~\mu$m) of the water-ice feature.

Despite their quantitative differences, a similar wavelength dependence can be seen for the two wavelengths. For small grains ($\Delta\phi\ll1$), the polarization degree is constant due to Rayleigh scattering. Once the phase shift exceeds unity, the polarization degree begins to decrease. However, as the grain radius increases, the polarization shows a weaker decrease and then begins to increase. Once the grain has a sufficient optical thickness for absorption ($\tauabs\sim10$), the polarization degree becomes almost constant again, corresponding to the reflection regime. 

There are two primary differences between the results for the two wavelengths. The first difference lies in the dust radius at which the reflection regime begins to dominate. This difference arises because the imaginary part of the refractive index at the center of the feature is approximately 45 times larger than that outside the feature. The second difference lies in the magnitude of the negative polarization. Compared with the outer wavelength of the feature, the central wavelength shows a relatively weak negative polarization. This trend occurs because, at the central wavelength, the trajectory in the $\Delta\phi-\tauabs$ plane moves from the single-scattering regime to the reflection regime without entering the multiple-scattering regime, as shown in Figure \ref{fig:regime}. As a result, negative polarization is not pronounced at the central wavelength.

\subsubsection{Dependence on the maximum grain radius}
Next, we consider distribution-averaged polarization properties. Figure \ref{fig:polsize} (middle and bottom) shows the distribution-averaged degree of polarization as a function of $\amax$ for a value of $\amin=0.005~\mu$m. 

The effect of the size distribution on the polarization feature can be assessed by a scaling argument. At the Rayleigh limit, $S_{11},~S_{12}\propto a^6$. In contrast, at the geometrical optics limit, these parameters are proportional to $a^2$. Therefore, the contribution of each grain to the integration of the numerator of Equation (\ref{eq:dist}) is
\begin{eqnarray}
S_{ij}n(a)a \propto \left\{ \begin{array}{ll}
a^{7-q} & (x\ll1), \\
a^{3-q} & (x\gg1). \\
\end{array} \right. \label{eq:scaling}
\end{eqnarray}
For the Rayleigh limit, the slope $7-q$ is positive, because we consider $q<7$. Thus, larger grains in the Rayleigh limit have a greater contribution to the integration. In contrast, at the geometrical optics limit, the sign of the slope $3-q$ changes depending on whether $q>3$ or $q<3$.

For $q>3$, the slope $3-q$ becomes negative. Thus, larger grains in the geometrical optics limit have a weak contribution to the integration. In this case, grains between the Rayleigh and geometrical optics limits, i.e., a few microns in size ($x\sim1-10$), will provide the dominant contribution to the integration. 

Because a size of a few microns is sufficient to reach the reflection regime at $\lambda=3.07~\mu$m, the distribution-averaged polarization remains high, whereas that at $\lambda=2.5~\mu$m decreases significantly (Figure \ref{fig:polsize} middle). Once $\amax$ enters the geometrical optics limit, the distribution-averaged polarization is almost independent of $\amax$.

For $q<3$, the slope $3-q$ becomes positive; thus, larger grains provide a greater contribution to the integration. As a consequence, the distribution-averaged polarization depends on $\amax$, as shown in Figure \ref{fig:polsize} (bottom). Because these large grains tend to be in the reflection regime, the polarization degree increases with $\amax$.

\begin{figure*}[t]
\begin{center}
\includegraphics[width=1.0\linewidth,keepaspectratio]{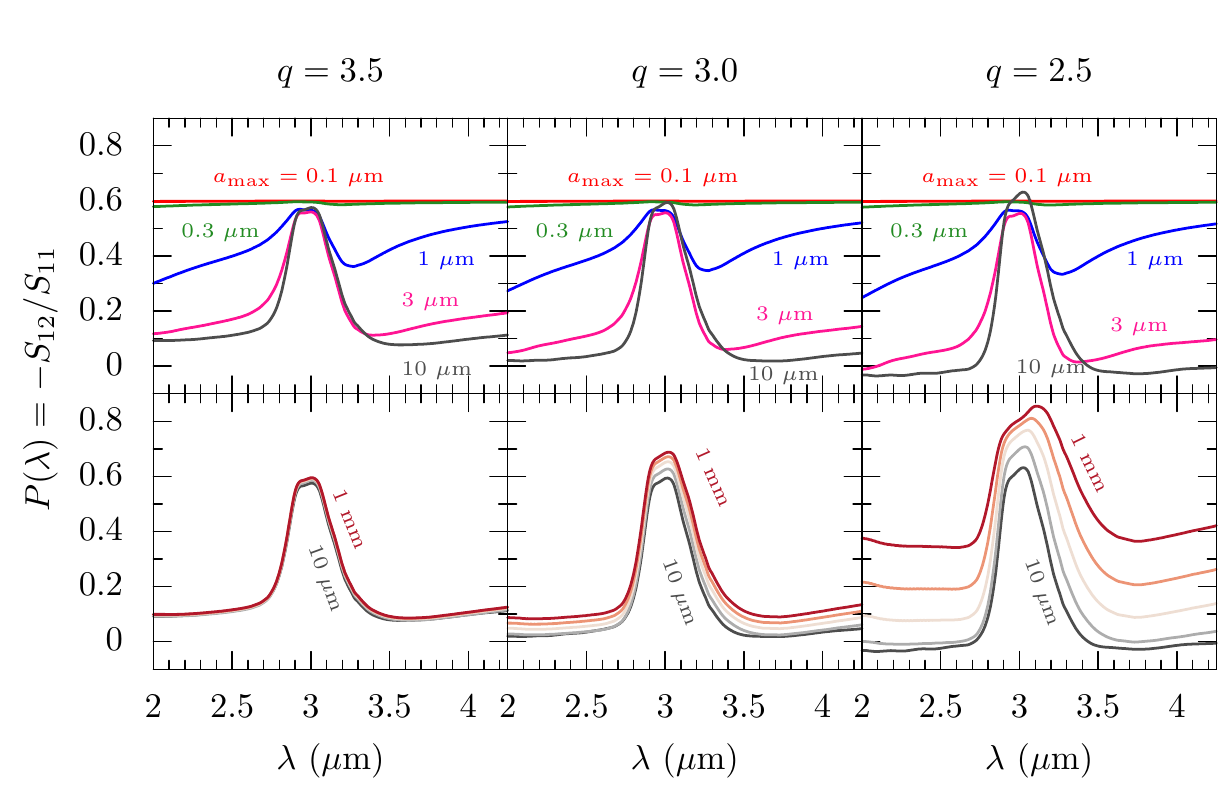}
\caption{A panoptic view of the scattering-polarization feature for various values of $\amax$ and $q$. The minimum grain radius is $0.005~\mu$m, and the scattering angle is fixed at $60^\circ$. The ice abundance is $\fice=0.3$. The top and bottom panels present results for a group of small $\amax$ values ($\amax=0.1~\mu$m, $0.3~\mu$m, 1$~\mu$m, $3~\mu$m, and $10~\mu$m) and a group of large $\amax$ values ($\amax=10~\mu$m, $30~\mu$m, $100~\mu$m, $300~\mu$m, and $1$ mm), respectively. From left to right, $q=3.5$, $3.0$, and $2.5$, respectively.}
\label{fig:sizedist1}
\end{center}
\end{figure*}

\begin{figure*}[t]
\begin{center}
\includegraphics[width=1.0\linewidth,keepaspectratio]{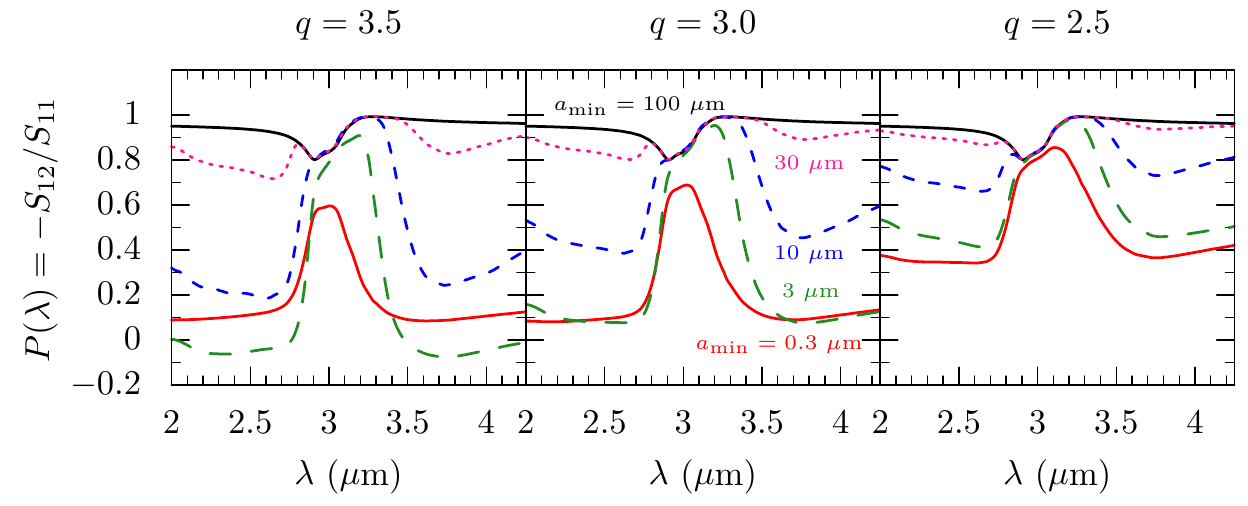}
\caption{Polarization feature for various values of $\amin$, with the maximum grain radius and scattering angle fixed at $\amax=1$ mm and $\theta=60^\circ$, respectively. The ice abundance is $\fice=0.3$. Each line corresponds to different values of $\amin$: $\amin=0.3~\mu$m (red solid), $3~\mu$m (green long-dashed), $10~\mu$m (blue short-dashed), $30~\mu$m (pink dotted), and $100~\mu$m (black solid). From left to right, $q=3.5$, $3.0$, and $2.5$, respectively.}
\label{fig:sizedist2}
\end{center}
\end{figure*}

Figure \ref{fig:sizedist1} shows a panoptic view of polarization features for various values of $\amax$ and $q$. Here, $\amax$ ranges from $0.1~\mu$m to $10~\mu$m (top panel) and from $10~\mu$m to $1$ mm (bottom panel). The former case is more likely to arise in interstellar media, molecular clouds, and protoplanetary disks, whereas the latter corresponds to debris disks.

For $\amax\lesssim10~\mu$m, the polarization feature is sensitive to $\amax$ regardless of the value of $q$. 
As $\amax$ increases, the polarization feature becomes stronger.
The feature tends to be pronounced for smaller $q$ values as the contribution of large grains increases. 
For $\amax\gtrsim10~\mu$m, the polarization feature for $q>3$ remains nearly constant. However, for $q<3$, the polarization degree increases for all ice band wavelengths. This result occurs because, even outside the feature, the grains become optically thick, and consequently, surface reflection begins to dominate (see also Figure \ref{fig:polsize}c). 

\subsubsection{Dependence on minimum grain radius} \label{sec:amin}
In debris disks, the minimum grain radius $\amin$ can reach tens of microns \citep[e.g.,][]{Mittal15}. Here, we study how such large $\amin$ affects the polarization feature.

Equation (\ref{eq:scaling}) indicates that the polarization feature is insensitive to the minimum grain radius as long as $\amin\ll\amax$ and $\amin\ll\lambda/(2\pi)$. Once the size parameter of the minimum grain approaches unity, i.e., $\amin\sim\lambda/(2\pi)\sim0.5~\mu$m at $\lambda=3~\mu$m, $\amin$ begins to influence the polarization feature.

Figure \ref{fig:sizedist2} shows the polarization feature for $\amin$ values of $0.3~\mu$m to $30~\mu$m, where the maximum grain radius is fixed at 1 mm. 
For all $q$ values considered, the polarization degree increases as $\amin$ increases for all ice band wavelengths. This behavior occurs because the fraction of grains in the reflection regime increases as $\amin$ increases. 
For $\amin=100~\mu$m, the wavelength dependence of the polarization is almost fully determined by surface reflection.
The degree of polarization near $\lambda\sim3~\mu$m fluctuates due to variations in the Brewster angle caused by the wavelength dependence of the refractive index (Figure \ref{fig:pwavel}).

For $\amax\gg10~\mu$m and $q>3$, the polarization feature is insensitive to $\amax$; therefore, this feature can be useful for inferring the minimum grain radius, for example, when studying dust in debris disks (Section \ref{sec:debris}).
However, for $q<3$, the polarization feature depends on both $\amin$ and $\amax$. Hence, in this case, determining the minimum and maximum grain radii solely from the observed polarization feature might not be straightforward. 

\subsection{Dependence on scattering angle} \label{sec:angle}

\begin{figure}[t]
\begin{center}
\includegraphics[height=6.0cm,keepaspectratio]{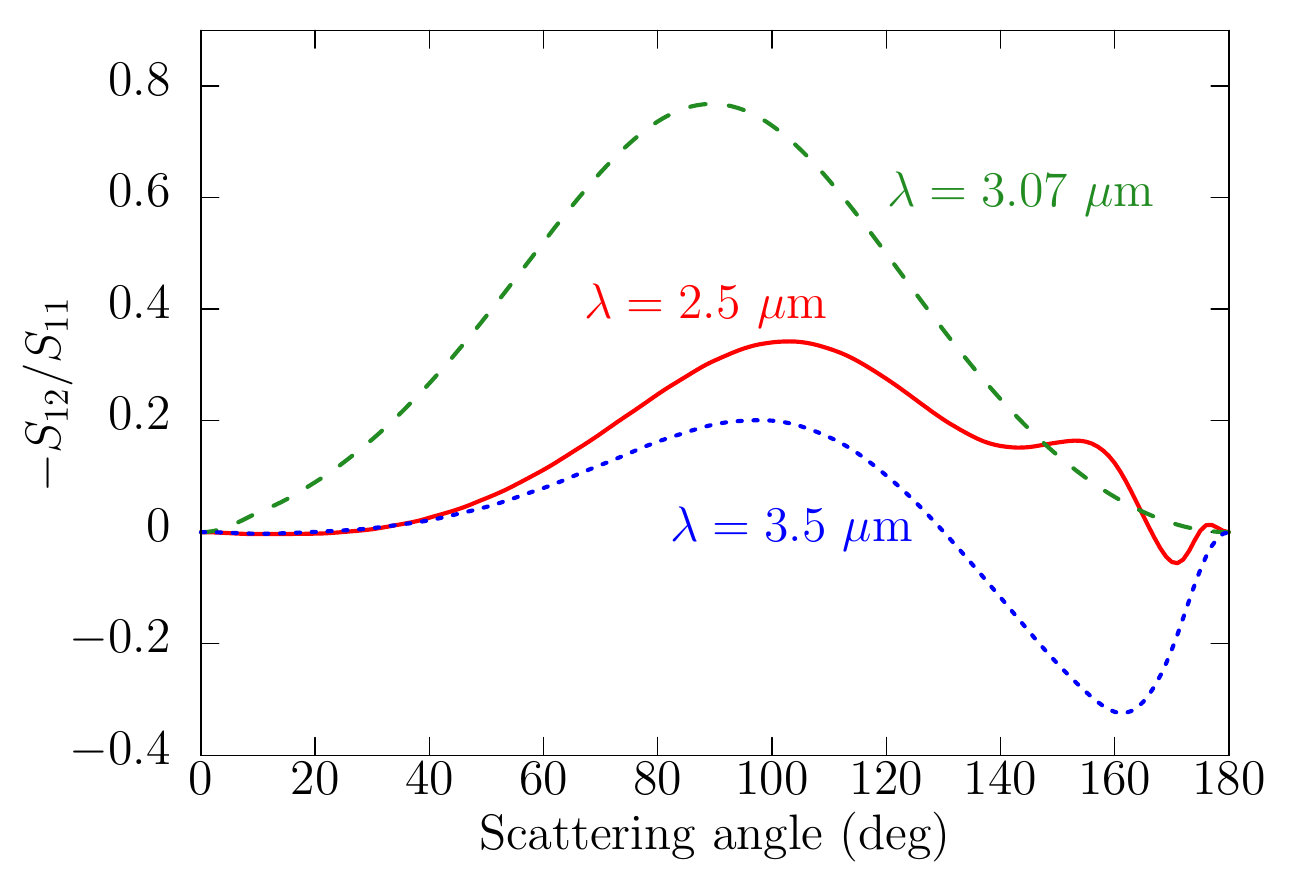}
\caption{Degree of polarization as a function of scattering angle for a size distribution of $\amin=0.005~\mu$m, $\amax=10~\mu$m, and $q=3.5$ and an ice abundance of $\fice=0.3$. The red, green, and blue lines show results for $\lambda=2.5~\mu$m, $3.07~\mu$m, and $3.5~\mu$m, respectively.}
\label{fig:polang}
\end{center}
\end{figure}

\begin{figure}[t]
\begin{center}
\includegraphics[height=6.0cm,keepaspectratio]{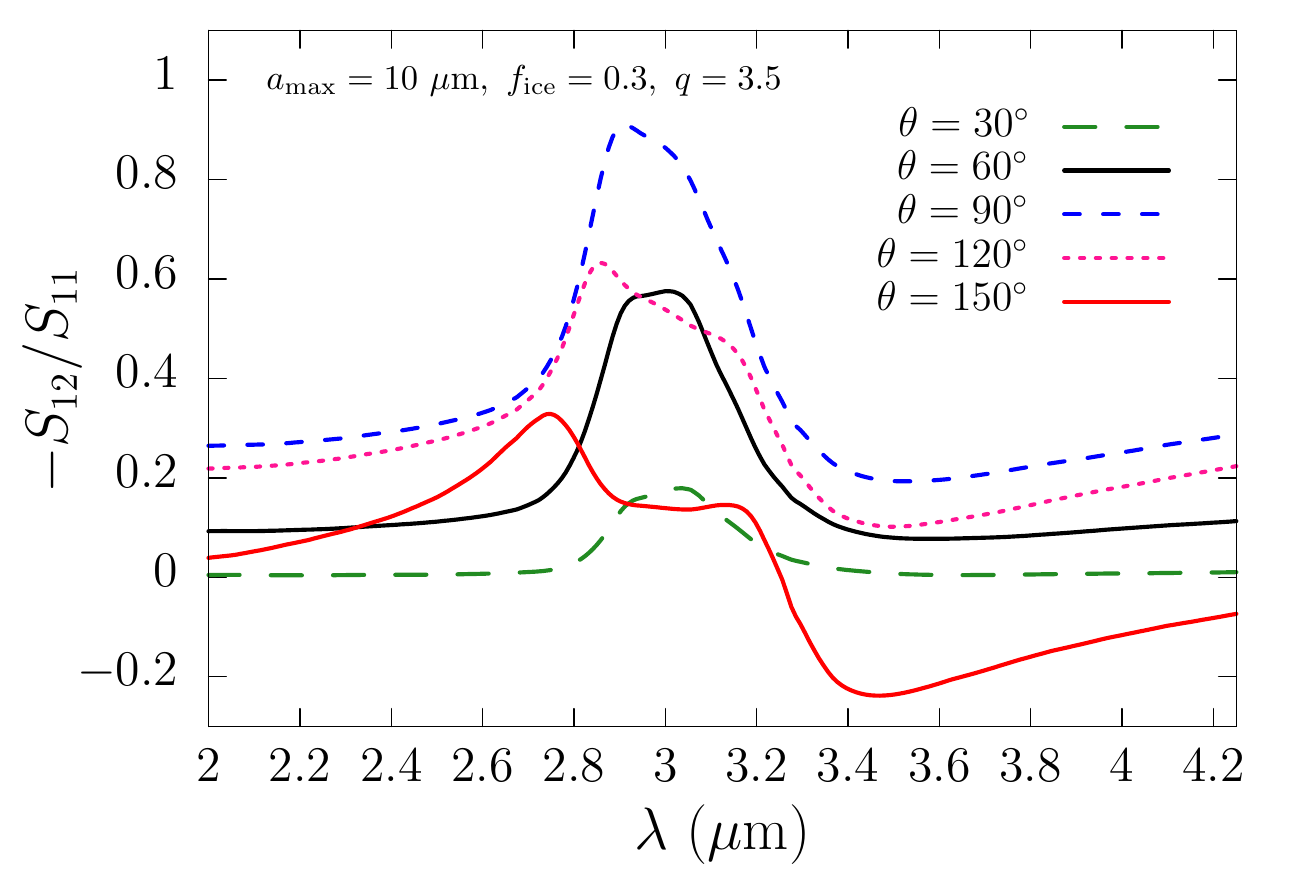}
\caption{Polarization degree as a function of wavelength for various scattering angles. The size distribution and abundance are the same as those in Figure \ref{fig:polang}.}
\label{fig:scatpol2}
\end{center}
\end{figure}

Scattering angles can differ greatly for each astrophysical situation. Here, we explore how the scattering angle affects the polarization feature. For $q=3.5$, the polarization feature is almost similar for $\amax\gtrsim3~\mu$m. Hence, we consider the case of $\amax=10~\mu$m as a representative case to show the angular dependence of the polarization feature. The results are shown in Figures \ref{fig:polang} and Figure \ref{fig:scatpol2}.

Figure \ref{fig:polang} shows that for $\lambda=3.07~\mu$m (the central wavelength of the ice feature), the angular dependence is almost symmetric with respect to $90^\circ$, with a maximal value of $\sim77\%$. For the outer two wavelengths of the feature, the absolute value of polarization tends to be lower at small scattering angles and higher at large scattering angles. Moreover, the back-scattering behavior differs between the outer two wavelengths of the feature. For example, a positive polarization hump occurs at $\sim153^\circ$ for $\lambda=2.5~\mu$m, but not for $\lambda=3.5~\mu$m. 

Figure \ref{fig:scatpol2} shows the wavelength dependence of the polarization degree for various scattering angles. For forward-side scattering angles ($\theta\lesssim90^\circ$), the polarization feature tends to be symmetric with respect to $\lambda\sim3~\mu$m. 
The polarization degree is higher at the center of the feature and lower at the outside.
In addition, the feature tends to be stronger as the scattering angle approaches $90^\circ$.
These tendencies are consistent with the findings presented in \citet{Pendleton90}.

For backward-side scattering angles ($\theta\gtrsim90^\circ$), the polarization feature becomes asymmetric. Here, the polarization degree in the outer region of the feature is higher than that at the center. This asymmetry is primarily caused by the enhanced polarization degree at the short-wavelength side of the feature, which can be attributed to the hump at $\theta\sim153^\circ$ for the case of $\lambda=2.5~\mu$m, as shown in Figure \ref{fig:polang}. 

The origin of the polarization hump corresponds to scattered light experiencing one internal reflection (see $p=2$ for Figure \ref{fig:pic}). The hump will occur when the internally reflected light is not absorbed and the internal reflection occurs near the Brewster angle. Therefore, the polarization hump represents essentially the same physics as a strongly polarized primary rainbow \citep[e.g.,][]{vdh57}. For a distribution of non-absorbing spheres, these two conditions tend to be satisfied when $1.1\lesssim n \lesssim 1.4$. When $n\gtrsim1.4$, the humps become less pronounced, and negative polarization starts to develop at back-scattering angles.

For the case of $\fice=0.3$, the real parts of the refractive index at the short- and long-wavelength sides of the feature are typically approximately $n\lesssim1.4$ and $n\gtrsim1.4$, respectively (see Figure \ref{fig:opcont}). Thus, the short-wavelength side tends to show a high degree of polarization due to the hump, whereas the long-wavelength side shows negative polarization. As a result, the polarization feature becomes asymmetric with respect to the central wavelength at back-scattering angles.

\subsection{Dependence on water-ice abundance} \label{sec:fice}
\begin{figure*}[t]
\begin{center}
\includegraphics[width=0.49\linewidth,keepaspectratio]{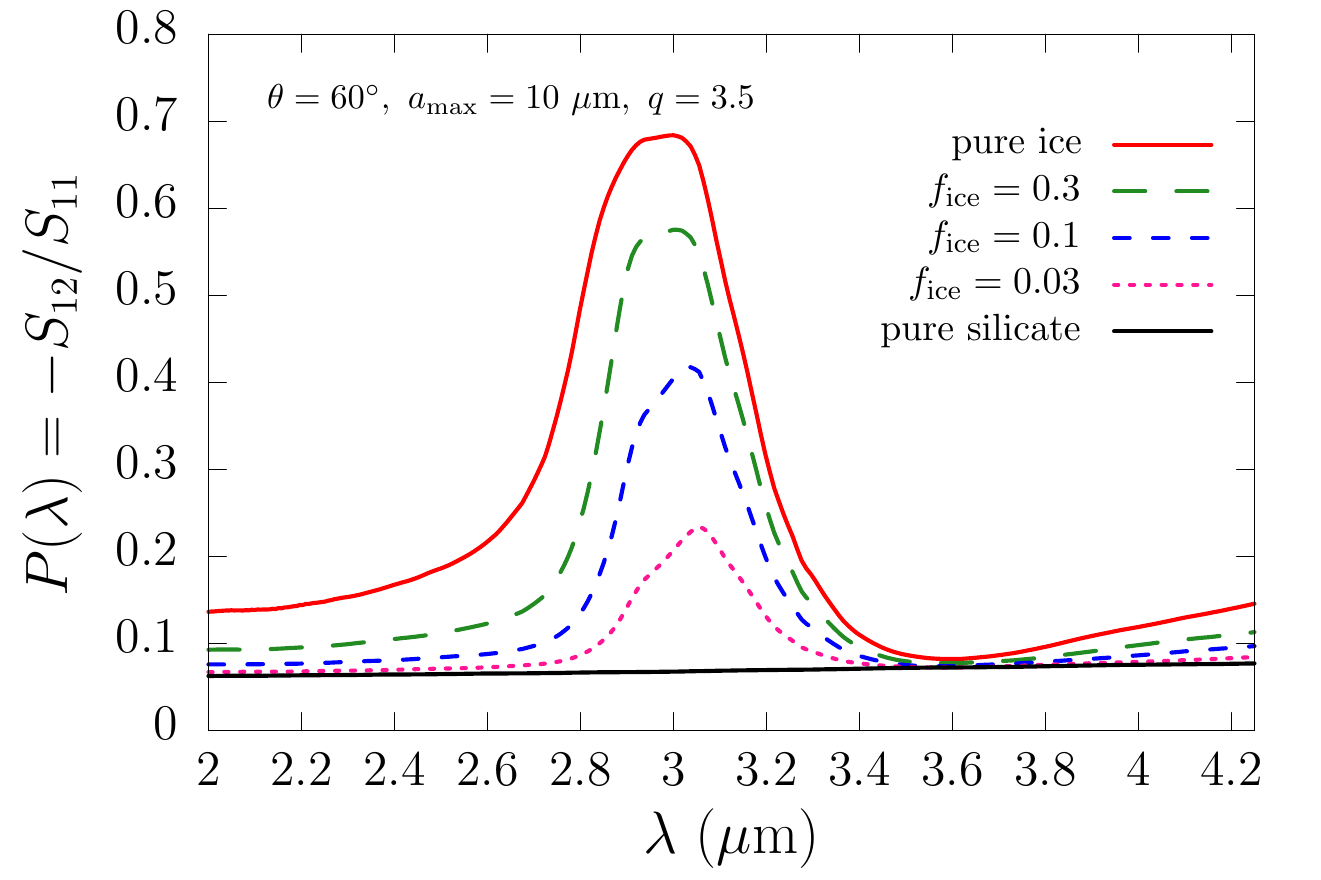}
\includegraphics[width=0.49\linewidth,keepaspectratio]{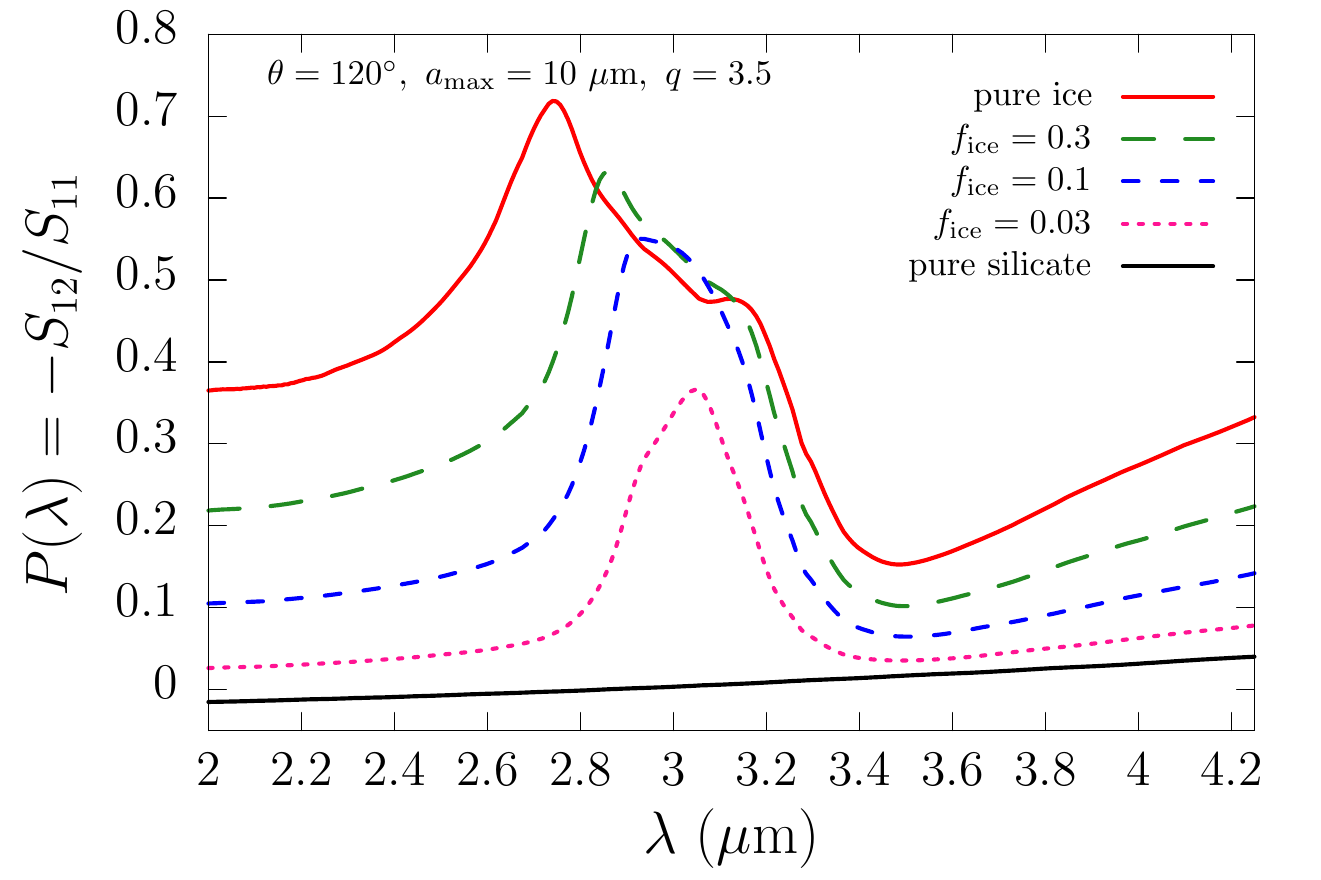}
\caption{Dependence of the polarization feature on the water-ice abundance. The left and right panels correspond to $\theta=60^\circ$ and $120^\circ$, respectively. The red and black solid lines show results for pure ice and pure silicate grains, whereas the long-dashed, short-dashed, and dotted lines show results for a mixture of ice and silicate with $\fice=0.3$, $0.1$, and $0.03$, respectively. }
\label{fig:polfice}
\end{center}
\end{figure*}

Here, we investigate how the ice abundance affects the polarization feature.
Figure \ref{fig:polfice} shows the polarization feature for various values of $\fice$.

At $\theta=60^\circ$, the polarization feature becomes more pronounced as the ice abundance increases. Even as $\fice$ varies, the profile of the feature remains almost symmetric with respect to $\lambda\sim3~\mu$m.

At $\theta=120^\circ$, the profile of the polarization feature varies significantly with $\fice$. The enhancement of the polarization degree at the short-wavelength side of the feature is sensitive to $\fice$, as it is closely related to the real part of the refractive index.
As $\fice$ decreases, the real part of the refractive index increases (see Figure \ref{fig:opcont}). If $\fice$ is lower than $\sim0.1$, the real part of the refractive index exceeds 1.4 for most wavelengths, and consequently, polarization enhancement no longer occurs at the short-wavelength side. As a result, the polarization feature tends to be symmetric with respect to $\lambda\sim3~\mu$m for ice-poor cases ($\fice\lesssim0.1$).

\section{Origin of the polarization feature for large icy grains} \label{sec:origin}
In Section \ref{sec:scatpol}, we mentioned that surface reflection is essential for the appearance of the polarization feature for large icy grains. In this section, we present a detail analysis of light scattered by large icy grains via the Debye series \citep{vdp37a, vdp37b, Hovenac92}. 
By using the Debye series, we can attain clear physical insights into the effect of surface reflection on polarization (Section \ref{sec:a10}). We can also determine how the polarization feature depends on the scattering angle (Section \ref{sec:debyeangle}).

\begin{figure}[t]
\begin{center}
\includegraphics[width=1.0\linewidth,keepaspectratio]{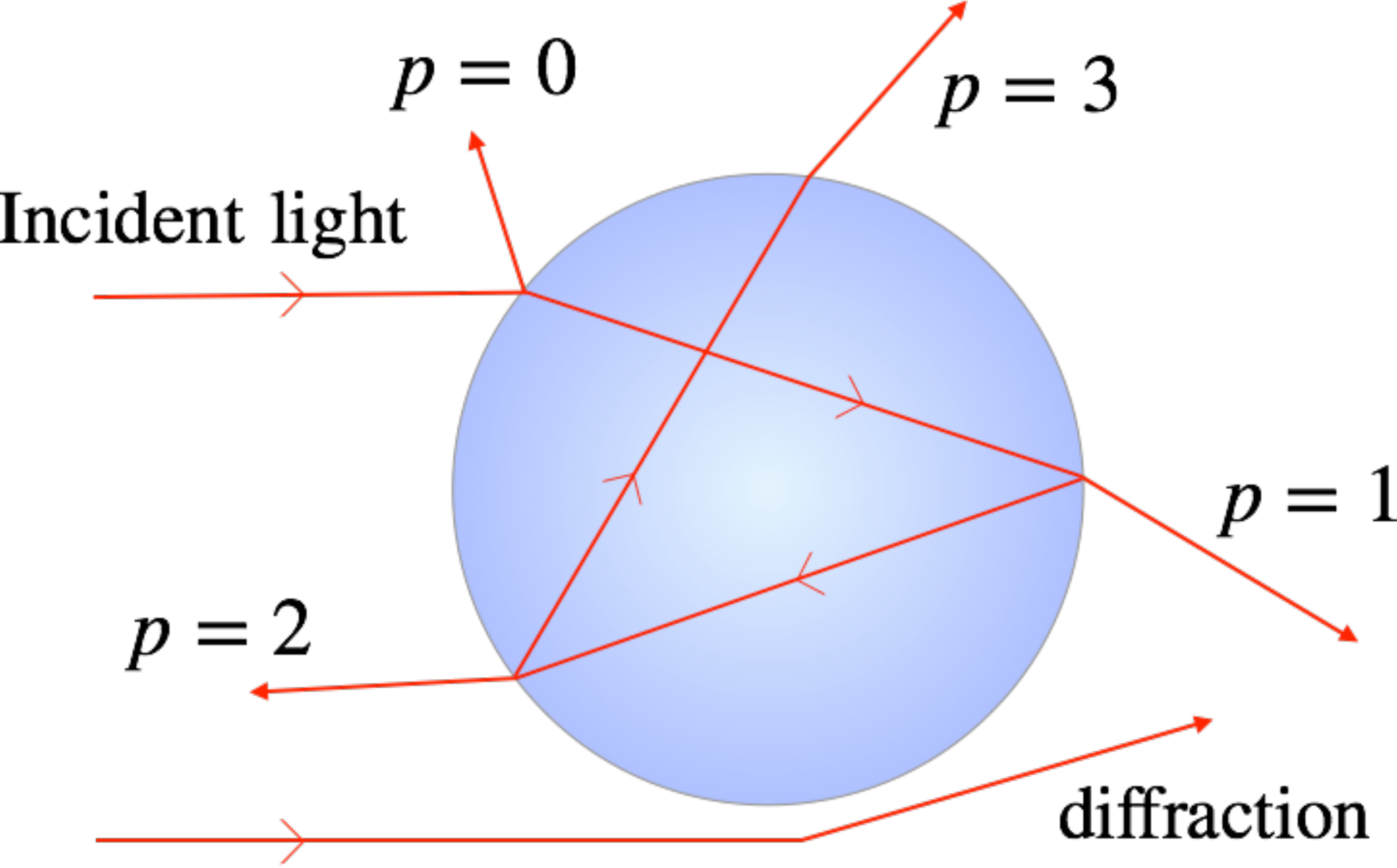}
\caption{Basic concept of the Debye series. The Debye series is a geometric series expansion of scattering coefficients with respect to the reflection coefficient. The scattered light is decomposed into diffracted light, surface-reflected light ($p=0$), twice-refracted light without internal reflection ($p=1$), and light subject to $p-1$ internal reflections ($p\ge2$).}
\label{fig:pic}
\end{center}
\end{figure}

The Debye series is a light-scattering solution for a homogeneous sphere and is an exact alternative of the Mie theory \citep{vdp37a, vdp37b, Hovenac92}. 
The primary advantage of the Debye series over the Mie theory is that it gives clear physical insights into light-scattering phenomena. 
A basic concept of the Debye series is illustrated in Figure \ref{fig:pic}. In Debye series calculations, we can decompose scattering matrix elements into a number of components in a rigorous manner: diffracted light, surface-reflected light ($p=0$), twice-refracted light without internal reflection ($p=1$), and light subject to $p-1$ internal reflection ($p\ge2$). In the following, we refer to the sum of diffraction and surface reflection as surface scattering. The formulation and benchmark calculations of the Debye series are presented in Appendix \ref{sec:debye}.

\subsection{Roles for surface and internal scattering in polarization} \label{sec:a10}
\begin{figure}[tbp]
\begin{center}
\includegraphics[width=1.0\linewidth,keepaspectratio]{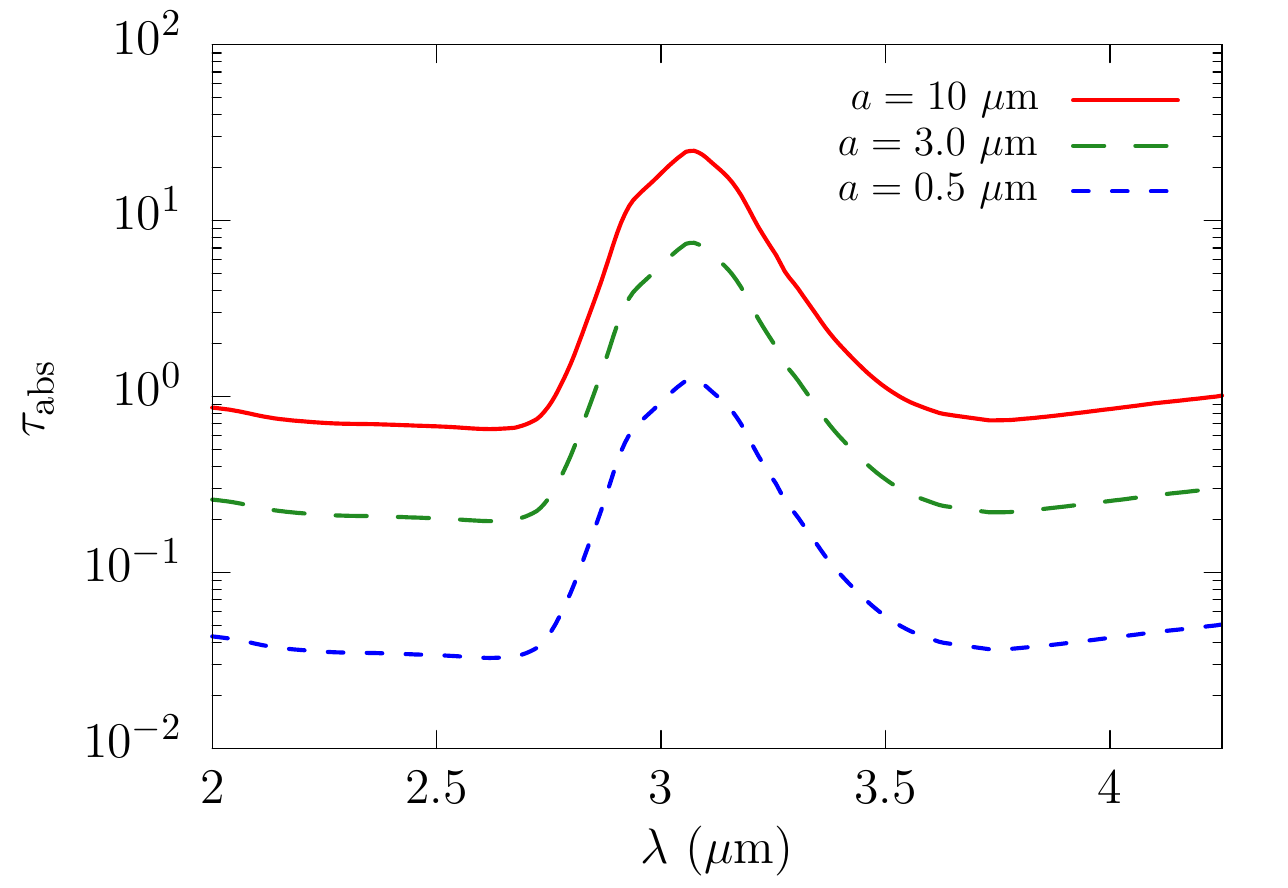}
\caption{Absorption optical depth of single-sized grains with $\fice=0.3$. The red, green, and blue lines present results for $a=10~\mu$m, $3~\mu$m, and $0.5~\mu$m, respectively.}
\label{fig:tauabs}
\end{center}
\end{figure}

\begin{figure*}[t]
\begin{center}
\includegraphics[width=0.495\linewidth,keepaspectratio]{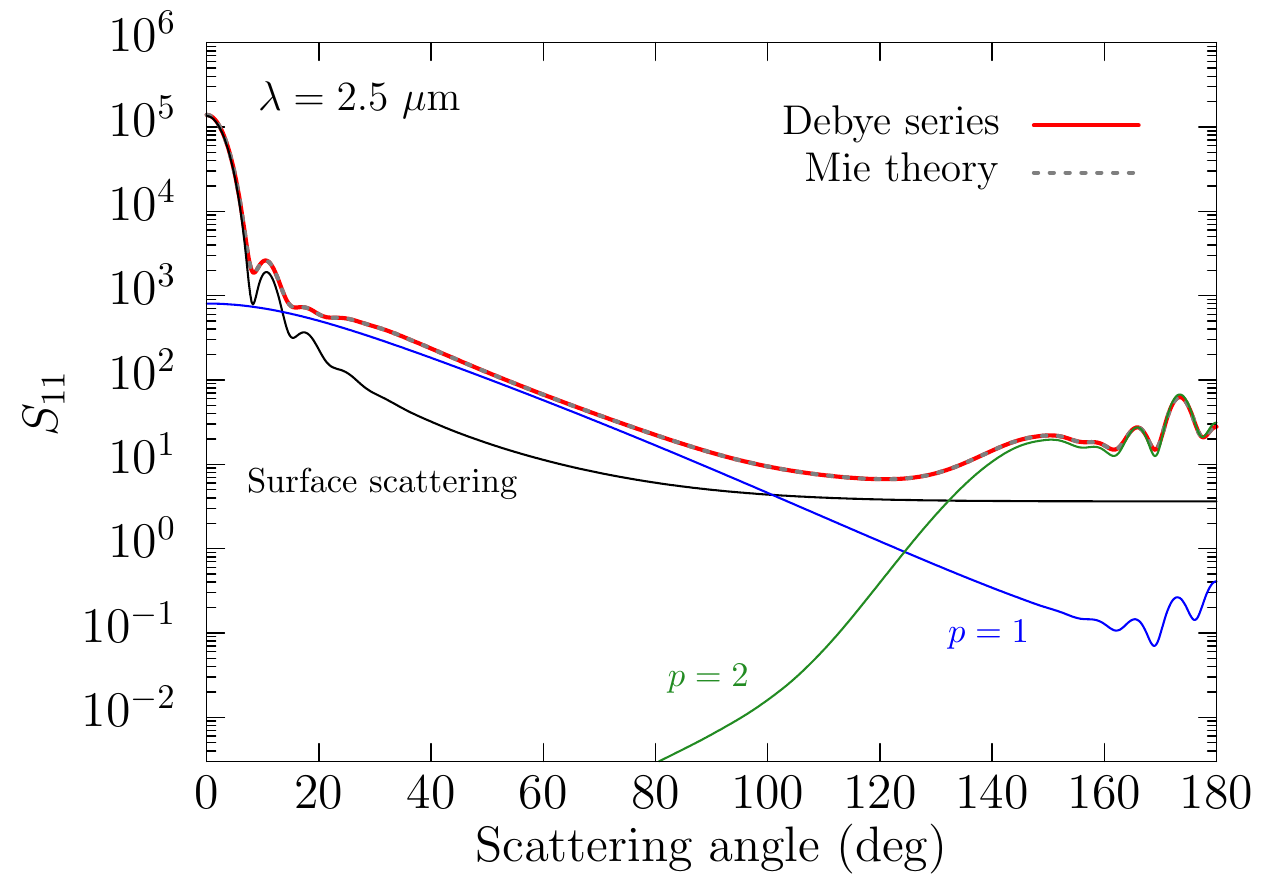}
\includegraphics[width=0.495\linewidth,keepaspectratio]{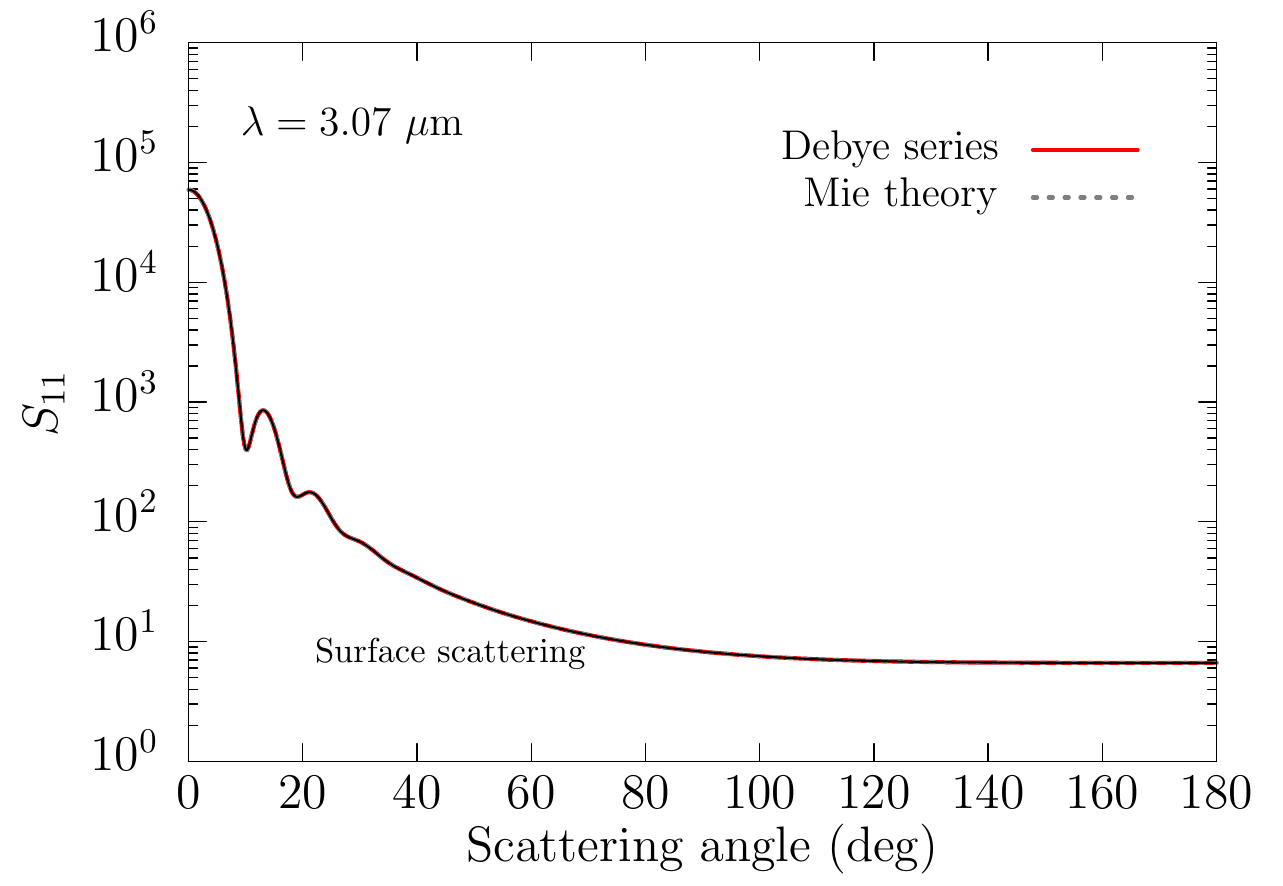}
\caption{$S_{11}$ for $\amean=10~\mu$m and $\fice=0.3$ as a function of scattering angle. The left and right panels show results for $\lambda=2.5~\mu$m (outside the feature) and $3.07~\mu$m (center of the feature), respectively. The red solid line presents the result obtained by the Debye series summed up to an infinite order ($p\to\infty$), whereas the gray dashed line presents the result obtained by the Mie theory. The results show excellent agreement, ensuring the validity of the Debye series calculations. Each term of the Debye series is also shown in each panel: surface scattering (diffraction plus $p=0$, black solid line), twice-refracted light ($p=1$, blue solid line), and light subject to one internal reflection ($p=2$, green solid line). Only surface scattering is shown in the right panel because the other components are negligibly small in this case.}
\label{fig:debyeS11}
\end{center}
\end{figure*}

Using the Debye series, we investigate the scattering properties of $10$-$\mu$m particles, which are optically thick near the central wavelengths of the feature but marginally thin at the outside wavelengths (see Figure \ref{fig:tauabs}). Thus, this case illustrates a typical polarization feature for $q=3.5$ and $\amax\gtrsim3~\mu$m.

Because the single-sized grains are larger than the wavelength, the optical properties show a strong oscillatory pattern.
Thus, we average the scattering matrix elements over a log-normal size distribution:
\begin{equation}
n(a)da \propto \exp\left[-\frac{(\ln(a/\amean))^2}{2\sigma^2}\right] d\ln{a},
\end{equation}
where the mean radius $\amean=10~\mu$m and $\sigma=0.1$. 

Figure \ref{fig:debyeS11} shows the scattering matrix element of the mean radius 10 $\mu$m at $\lambda=2.5~\mu$m (outside the feature) and $\lambda=3.07~\mu$m (center of the feature). 

The angular distribution of the scattered light intensity differs significantly for the two wavelengths.
At $\lambda=2.5~\mu$m, the scattered light primarily arises from surface-scattered light, twice-refracted light ($p=1$), and light subjected to one internal reflection ($p=2$). 
By contrast, at $\lambda=3.07~\mu$m, surface scattering is dominant at all scattering angles. This difference occurs because at $\lambda=3.07~\mu$m, the grains are highly absorbing; thus, the higher-order effects ($p\ge1$) are significantly attenuated. Therefore, the suppression of internally reflected light causes differences in the scattering properties of these two wavelengths.

\begin{figure*}[tbp]
\begin{center}
\includegraphics[width=0.49\linewidth,keepaspectratio]{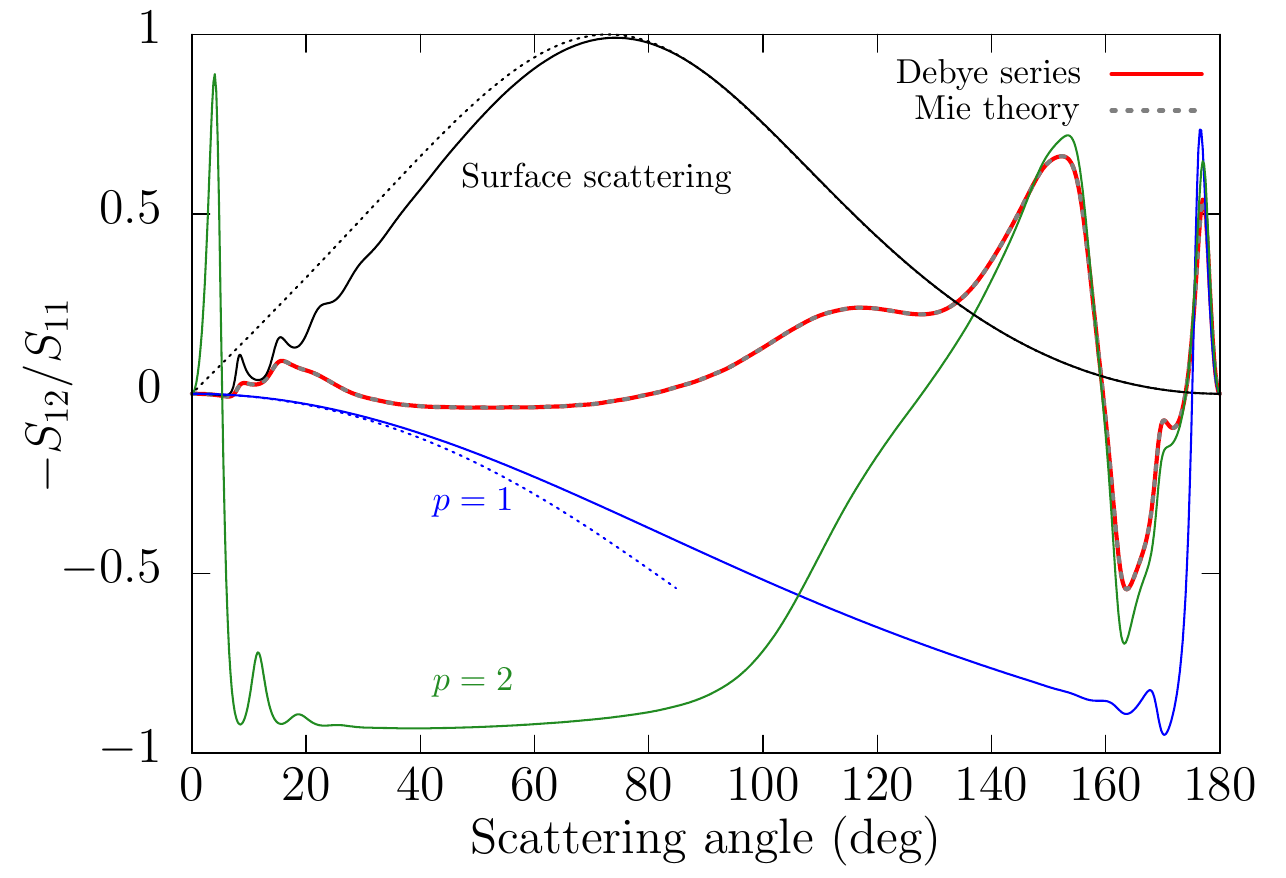}
\includegraphics[width=0.49\linewidth,keepaspectratio]{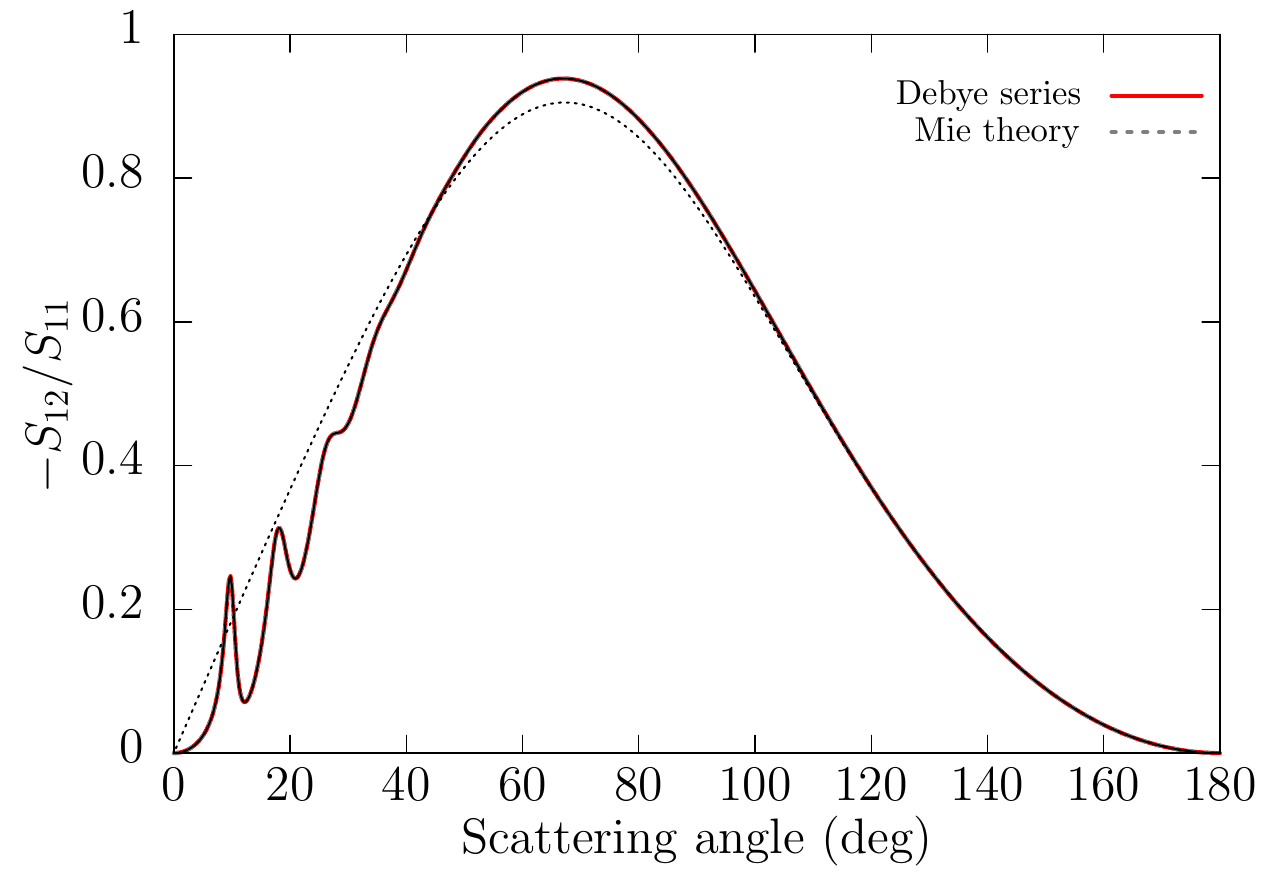}
\caption{Same as Figure \ref{fig:debyeS11}, but for the degree of polarization.
The black and blue dotted lines present solutions for Equations (\ref{eq:Pp0}) and (\ref{eq:Pp1}), respectively.}
\label{fig:debyePOL}
\end{center}
\end{figure*}

Figures \ref{fig:debyePOL} compares the angular dependence of the polarization degree at the two wavelengths. In each panel, we show the degree of polarization for each scattered light component. 
At $\lambda=2.5~\mu$m, the polarization degree at forward-scattering angles is low because highly polarized surface-scattered light is significantly depolarized by twice-refracted light, which is negatively polarized ($p=1$). At back-scattering angles, the scattered light with $p=2$ tends to determine the polarization degree.
In contrast, at $\lambda=3.07~\mu$m, the polarization degree is almost fully determined by that of surface scattering.

\subsection{Interpreting the polarization properties with the Fresnel reflection formulas}
The polarization degree of surface scattering can be understood by the Fresnel formula, which is derived for $a\to\infty$.
The polarization degree of Fresnel reflection is given by \citep{Bohren83}:
\begin{eqnarray}
P_{p=0}(\theta)=\frac{R_\perp-R_{||}}{R_\perp+R_{||}}, \label{eq:Pp0}
\end{eqnarray}
where $R_\perp$ and $R_{||}$ are the reflectances for the electric field vectors perpendicular and parallel to the plane of incidence, respectively. The reflectances are defined by
\begin{eqnarray}
R_\perp=\left|\frac{\cos\Theta_i-m\cos\Theta_t}{\cos\Theta_i+m\cos\Theta_t}\right|^2,~
R_{||}=\left|\frac{\cos\Theta_t-m\cos\Theta_i}{\cos\Theta_t+m\cos\Theta_i}\right|^2,\nonumber\\
\end{eqnarray}
where $\Theta_i$ and $\Theta_t$ are the angles between the wavenumber vectors of incident and refracted light, respectively, and the normal vector to the grain surface. $\Theta_i$ is defined in the range $[0,\pi/2]$, and the refraction angle is given by Snell's law: $m\sin\Theta_t=\sin\Theta_i$. For a geometric ray with $p=0$, the scattering angle $\theta$ and $\Theta_i$ are related via $\theta=\pi-2\Theta_i$. 

The polarization obtained by Equation (\ref{eq:Pp0}) is shown in Figure \ref{fig:debyePOL}. 
Although the mean radius of $10~\mu$m is not sufficiently large for the Fresnel formulas to provide a good approximation, the results highlight an important point. The externally reflected light will be almost perfectly polarized at the scattering angle $\theta=\pi-2\Theta_p$, where $\tan\Theta_p=n$ is the Brewster angle. For example, $m=1.33$ (non-absorbing) gives $\Theta_p\simeq53^\circ$ and $\theta\simeq74^\circ$. As a result, the surface-reflected light will be highly polarized at scattering angles close to the Brewster angle. At $\lambda=2.5~\mu$m, scattered light with $p=0$ shows a polarization degree of $94\%$ at $\theta=60^\circ$.

For twice-refracted light ($p=1$), the degree of polarization is characterized by the transmission coefficients of the Fresnel formulas:
\begin{eqnarray}
P_{p=1}(\theta)=\frac{T_\perp^{\mathrm{in}}T_\perp^{\mathrm{out}}-T_{||}^{\mathrm{in}}T_{||}^{\mathrm{out}}}{T_\perp^{\mathrm{in}}T_\perp^{\mathrm{out}}+T_{||}^{\mathrm{in}}T_{||}^{\mathrm{out}}}, \label{eq:Pp1}
\end{eqnarray}
where $T^{\mathrm{in}}$ and $T^{\mathrm{out}}$ represent the transmittances for the first and second refraction, defined by
\begin{eqnarray}
T_\perp^{\mathrm{in}}&=&\left|\frac{2\cos\Theta_i}{\cos\Theta_i+m\cos\Theta_t}\right|^2,~
T_{||}^{\mathrm{in}}=\left|\frac{2\cos\Theta_i}{\cos\Theta_t+m\cos\Theta_i}\right|^2,\nonumber
\\
T_\perp^{\mathrm{out}}&=&\left|\frac{2m\cos\Theta_t}{\cos\Theta_i+m\cos\Theta_t}\right|^2,~
T_{||}^{\mathrm{out}}=\left|\frac{2m\cos\Theta_t}{\cos\Theta_t+m\cos\Theta_i}\right|^2.\nonumber\\
\end{eqnarray}
For a geometrical ray with $p=1$, the scattering angle is given by $\theta=2(\Theta_i-\Theta_t)$. For $m=1.33$ (non-absorbing), the scattering angle varies from $0^\circ$ to $82.5^\circ$ as $\Theta_i$ varies from 0 to $\pi/2$. 
A plot of Equation (\ref{eq:Pp1}) is shown in Figure \ref{fig:debyePOL}. In contrast to the case of $p=0$, scattered light with $p=1$ is weakly and negatively polarized. At $\lambda=2.5~\mu$m, the polarization degree of scattered light with $p=1$ is approximately $-28\%$ at $\theta=60^\circ$.

For $\lambda=2.5~\mu$m, the polarization degree has a hump at $\theta\sim152^\circ$, which is attributed to the scattered light component of $p=2$. 
The scattering angle at which scattered light with $p=2$ tends to be dominant can be estimated by the primary rainbow angle.
If we ignore surface waves \citep{Hovenac92}, a geometric ray of $p=2$ covers scattering angles of $\theta>\theta^{(\mathrm{R})}\equiv4\Theta_t^{(\mathrm{R})}-2\Theta_i^{(\mathrm{R})}$, where $\theta^{(\mathrm{R})}$ is the scattering angle of the primary rainbow. $\Theta_i^{(\mathrm{R})}$ and $\Theta_t^{(\mathrm{R})}$ are given by \citep{Bohren83}
\begin{eqnarray}
\cos\Theta_i^{(\mathrm{R})}&=&\sqrt{\frac{m^2-1}{3}}, \label{eq:rainbow}\\
m\sin\Theta_t^{(\mathrm{R})}&=&\sin\Theta_i^{(\mathrm{R})}.
\end{eqnarray}
For $m=1.33$, $\theta^{(\mathrm{R})}\simeq138^\circ$, and for $m=1.4$, $\theta^{(\mathrm{R})}\simeq147^\circ$. Therefore, the effect of $p=2$ primarily arises for back-scattering angles.

Scattered light with $p=2$ is highly polarized when internal reflection occurs at angles close to the Brewster angle, that is, when 
\begin{equation}
\tan\Theta_t=\frac{1}{n}. \label{eq:intb}
\end{equation} 
At the primary rainbow angle, Equation (\ref{eq:intb}) is satisfied when $m=\sqrt{2}\simeq1.4$, where we have assumed a non-absorbing sphere.
Therefore, the refractive index of water ice, particularly at the short-wavelength side of the 3-$\mu$m feature, is found to be favorable for strong polarization of scattered light with $p=2$ (see the left panel in Figure \ref{fig:opcont}).

\subsection{Wavelength dependence of the polarization degree} \label{sec:debyeangle}
\begin{figure}[tbp]
\begin{center}
\includegraphics[width=1.0\linewidth,keepaspectratio]{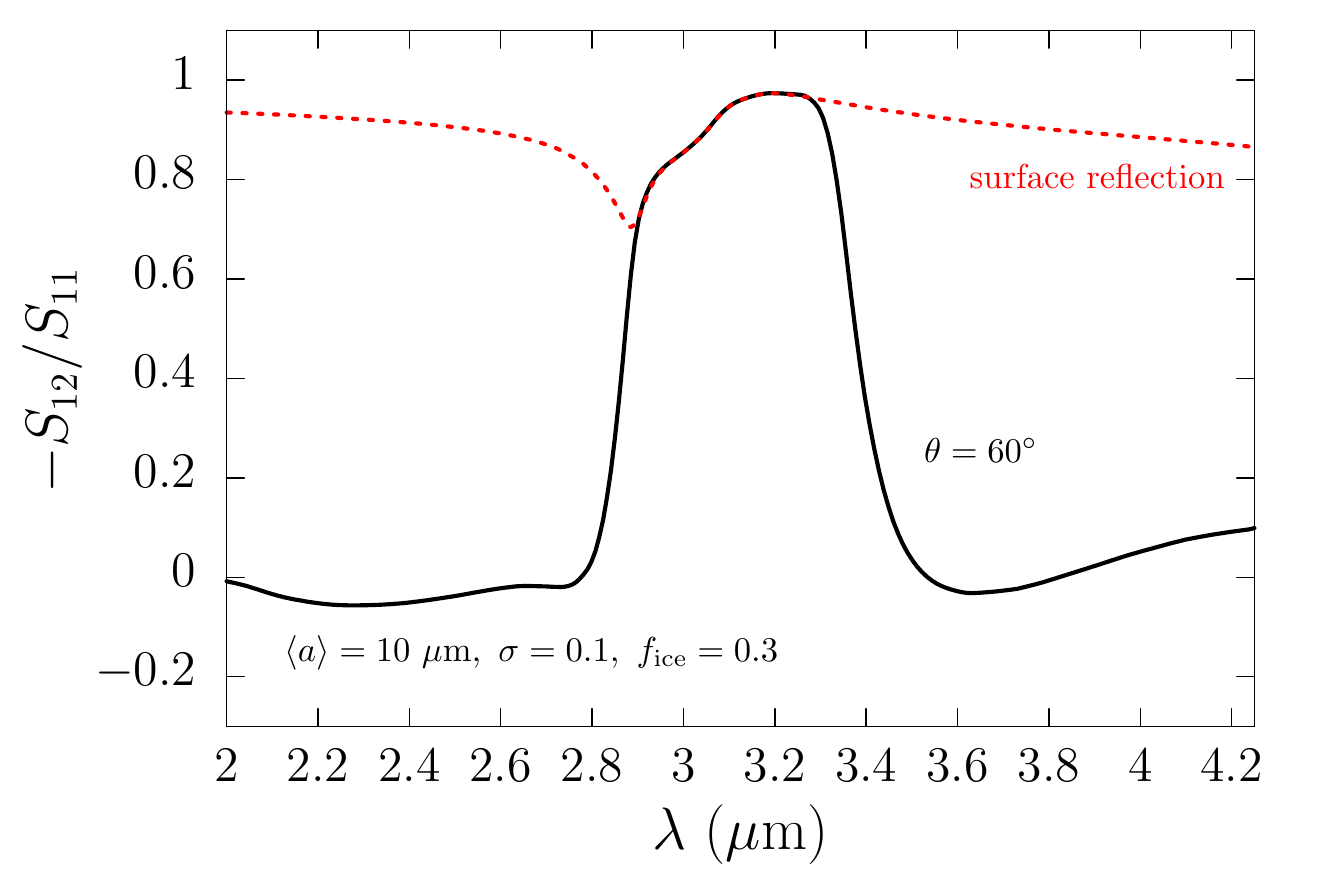}
\includegraphics[width=1.0\linewidth,keepaspectratio]{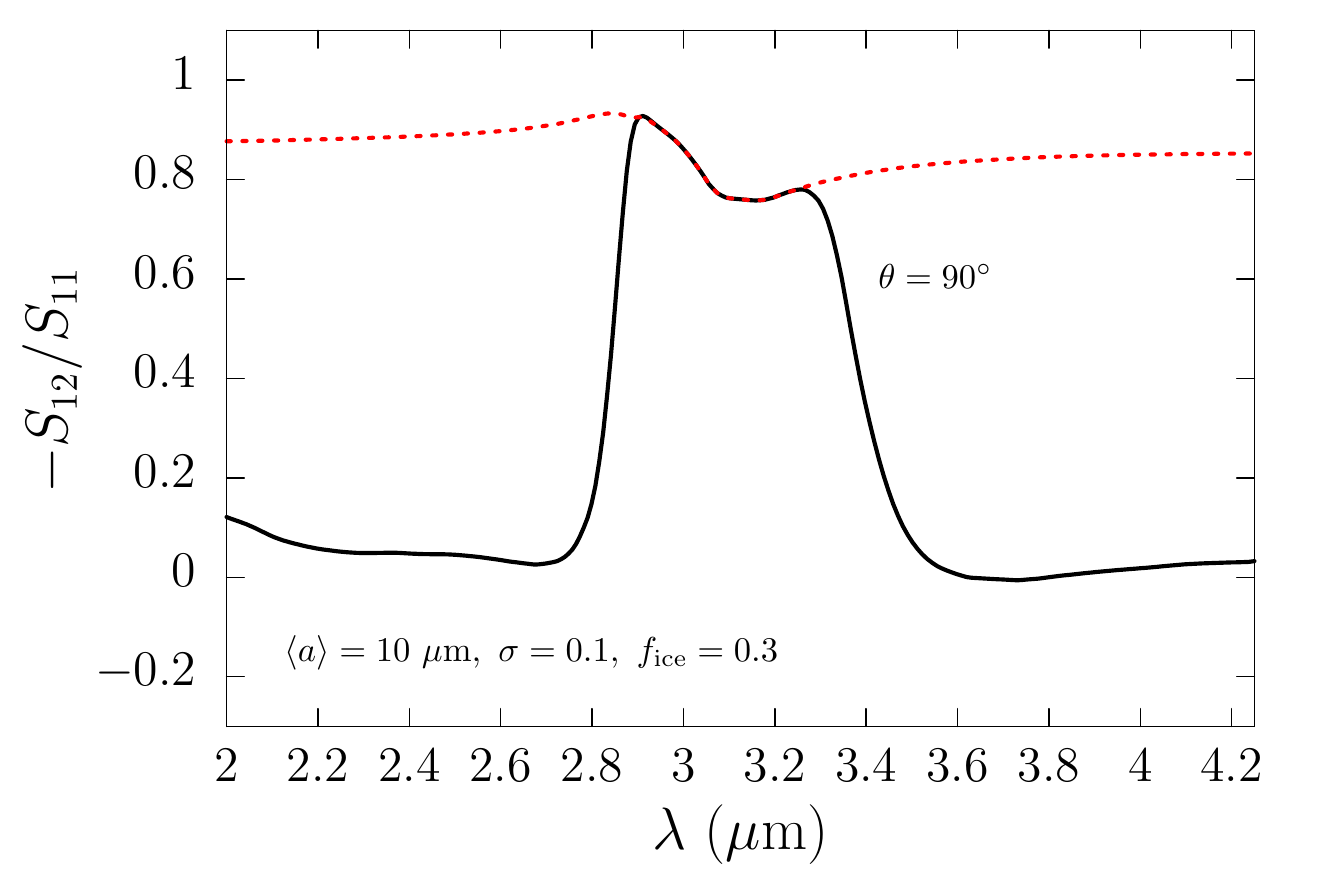}
\includegraphics[width=1.0\linewidth,keepaspectratio]{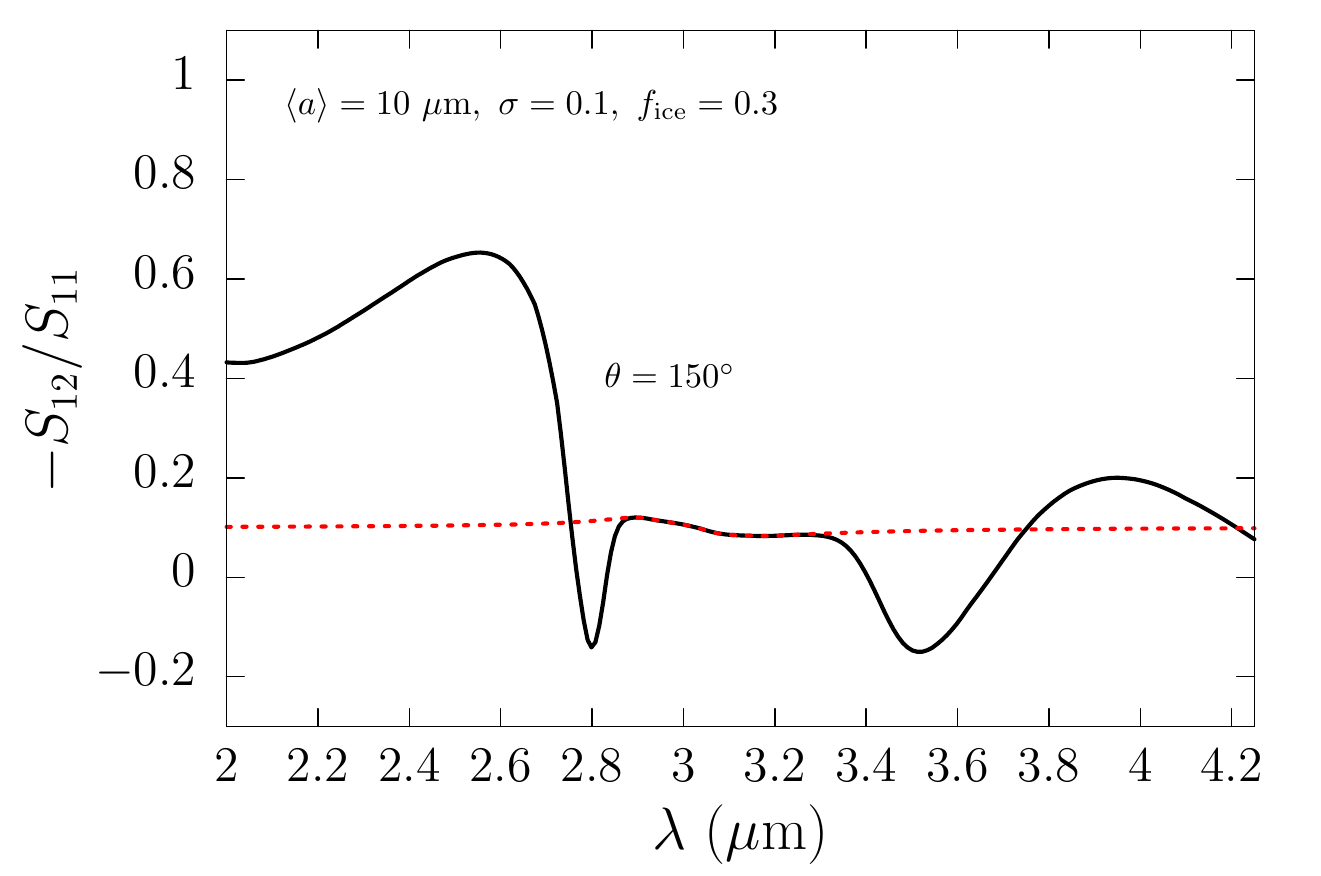}
\caption{Polarization degree as a function of wavelength for $\amean=10~\mu$m and $\fice=0.3$. The top, middle, and bottom panels correspond to scattering angles of $\theta=60^\circ$, $90^\circ$, and $150^\circ$, respectively.
The black solid line presents the full scattering solution (Mie theory), whereas the red dashed line presents results for surface scattering.}
\label{fig:pwavel}
\end{center}
\end{figure}

Figure \ref{fig:pwavel} shows the wavelength dependence of the polarization degree obtained by the Mie theory and the surface reflection ($p=0$) at scattering angles of $\theta=60^\circ$, $90^\circ$, and $150^\circ$. Figure \ref{fig:pwavel} clearly illustrates the influence of surface reflection on the polarization feature.

For all cases shown in Figure \ref{fig:pwavel}, the polarization degree at the feature center is characterized by that of the surface reflection. The polarization degree of the surface reflection depends on wavelength because the Brewster angle depends on the refractive index, which is determined by the wavelength. 
The polarization degree increases as the Brewster angle approaches a given scattering angle; conversely, the polarization decreases as the Brewster angle moves away from the scattering angle.
Because the real part of the refractive index varies rapidly near the feature center, the polarization degree fluctuates at these wavelengths. 
This kind of wavelength dependence can be observed for large values of $\amin$, as shown in Figure \ref{fig:sizedist2}. 

The symmetric and asymmetric polarization features discussed in Section \ref{sec:angle} are caused by a difference in the order of scattered waves dominating the scattering properties. 

A symmetric feature is generally observed for forward-side angles ($\theta\lesssim90^\circ$). At these angles, the scattering property will be described as either $p=0$ (for optically thick grains) or $p=1$ (for optically thin grains) (Figures \ref{fig:debyeS11} and \ref{fig:debyePOL}). 
Because light with $p=0$ and $p=1$ is strongly and weakly polarized, respectively, a clear contrast in the degree of polarization can be seen. In addition, the polarization degree for $p=1$ is less sensitive to the real part of the refractive index than that for $p=2$; hence, the polarization feature tends to be symmetric with respect to $\lambda\sim3~\mu$m.

The asymmetric feature can be seen at backward-side angles, where $p=0$ (optically thick grains) or $p=2$ (optically thin grains) dominate the scattering. In this case, light with $p=0$ or $p=2$ can be moderately polarized because both cases experience reflection. 
In particular, the polarization degree at the short-wavelength side is largely enhanced by internal Brewster scattering. 
As a consequence, the polarization feature becomes asymmetric with respect to $\lambda\sim3~\mu$m ($\theta=150^\circ$ in Figure \ref{fig:pwavel}).

To summarize, the polarization feature forms as follows. At the feature center, a highly absorbing refractive index renders the grains optically thick; consequently, the surface reflection, which is highly polarized near the Brewster angle, produces a high degree of polarization. In contrast, for the region outside the feature, internally transmitted/reflected light is dominant rather than surface scattering. In this case, the polarization degree deviates from that produced by surface reflection, resulting in either a symmetric or asymmetric polarization feature.

\section{Comparison with observations of a low-mass protostar envelope} \label{sec:obs}

\citet{Kobayashi99} reported the first detection of a polarization excess at $\sim3~\mu$m for a low-mass young stellar object: the envelope of the low-mass protostar L1551 IRS 5. However, the obtained polarization is inconsistent with the prediction by \citet{Pendleton90}. Here, we compare our models of scattering polarization with previous observations.

The detected near-infrared polarization is thought to arise from the cavity of the envelope evacuated by the jet \citep{Nagata83, Strom88, Kobayashi99}.
As both the inclination angle of the jet and the half-opening angle of the cavity are estimated to be $\sim45^\circ$ \citep{Pyo02}, the typical scattering angle is approximately $\sim90^\circ$.

To compare our models with observational data, we digitized the data points shown in Figure 3 in \citet{Kobayashi99} by using \texttt{PlotDigitizerX} software, as the original digital data are no longer available (N. Kobayashi 2020, private communication). Four data points (channels 8, 26, and 27 for PASP2 and channel 11 for PSP) shown in their Figure 3 are unreadable because their error bars are hidden by other foreground data points. For comparison, we also plotted the observation data tabulated in \citet{Nagata83}.

\begin{figure}[t]
\begin{center}
\includegraphics[height=6.0cm,keepaspectratio]{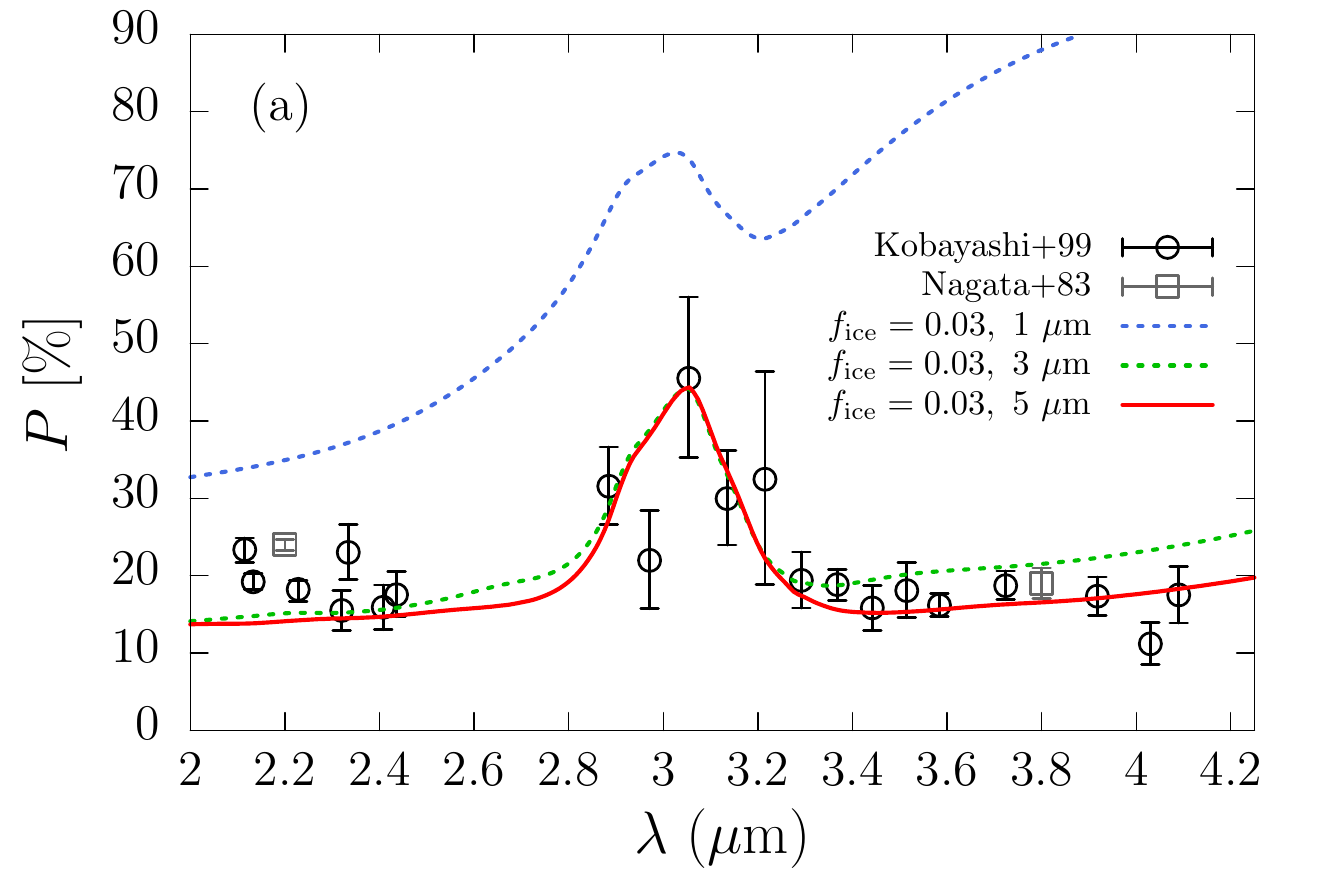}
\includegraphics[height=6.0cm,keepaspectratio]{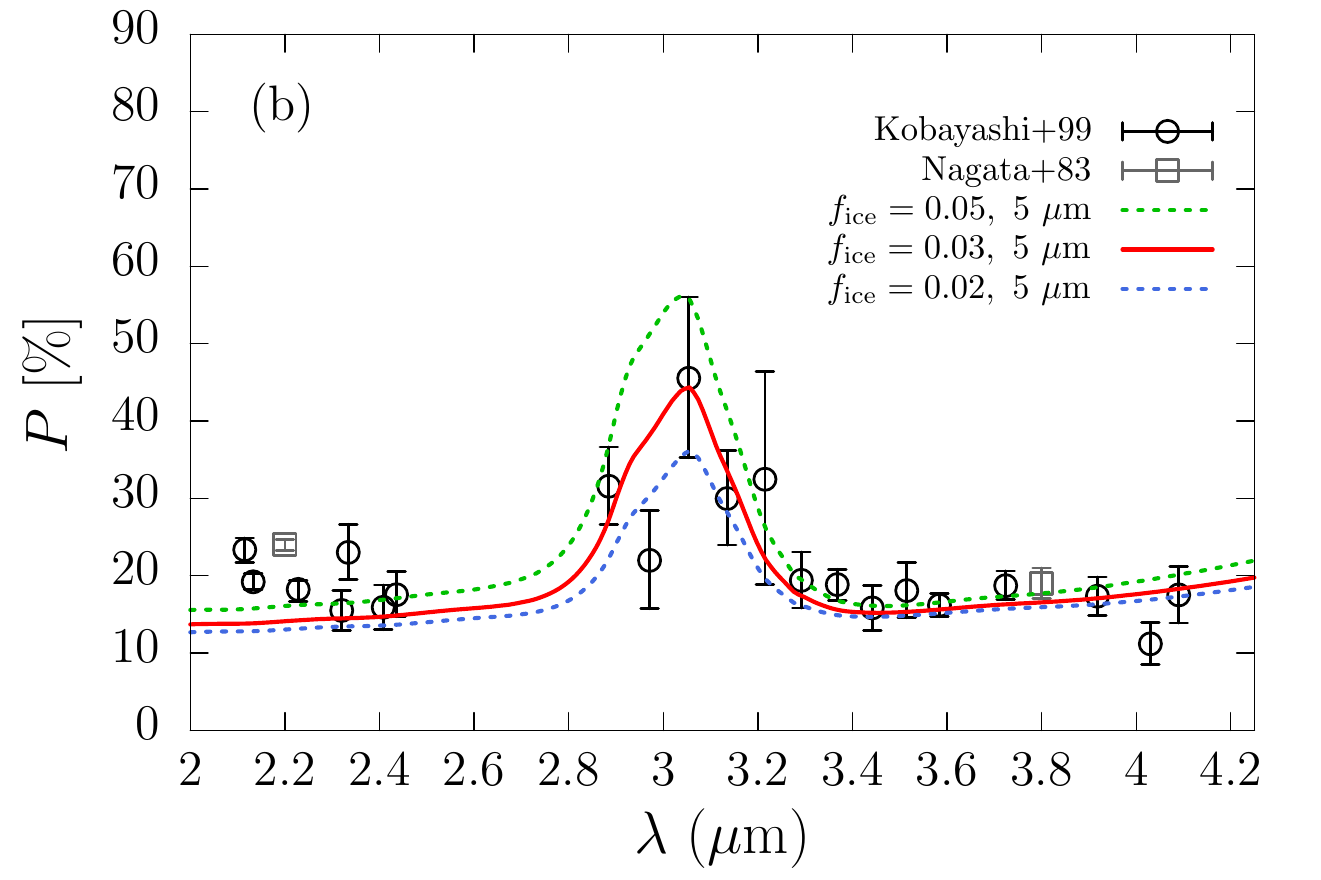}
\caption{Comparison with polarimetric observations of the protostar envelope L1551 IRS 5 \citep{Nagata83, Kobayashi99}. The scattering angle is assumed to be $90^\circ$. (a) The blue, green, and red lines show $\amax=1~\mu$m, $3~\mu$m, and $5~\mu$m, respectively. (b) Same as (a), but for different water-ice abundances. }
\label{fig:k99}
\end{center}
\end{figure}

Figure \ref{fig:k99} (a) compares the observations with our models for various values of $\amax$, while the ice abundance is fixed at $\fice=0.03$.
The observed polarization fraction is nearly constant outside the feature. 
For $\amax=1~\mu$m, the polarization decreases as the wavelength decreases \citep{Pendleton90}; hence, the model fails to reproduce the flat polarization.
The observations might be explained if $\amax\gtrsim3~\mu$m. A flat polarization curve is a characteristic of the polarization feature in the reflection regime. 

The excess polarization can potentially be used to assess the water-ice abundance. Figure \ref{fig:k99} (b) shows results for varying water-ice abundances. The observed polarization excess is consistent with models of $\fice\sim0.02 - 0.05$. The abundance inferred from our models is lower than the commonly assumed abundance \citep[e.g.,][]{Pollack94}. Therefore, our model suggests a depletion of water ice in the cavity region.

To summarize, our model suggests that the envelope of L1551 IRS 5 contains micron-sized ice-poor particles. The presence of $\mu$m-sized grains may indicate grain growth in the envelope region, which is also favored by the fact that envelopes often show coreshine \citep{Steinacker10,Pagani10}. The depletion of water ice suggests the necessity of ice disruption in the cavity region, e.g., due to photodesorption \citep{Oka12} or rotational disruption \citep{Hoang20, Tung20}.

\section{Implications for various astrophysical environments} \label{sec:disc}
As the identification of large icy grains can be important for various astronomical environments, we discuss some potential applications of the polarization feature.

\subsection{Missing-oxygen problem in diffuse molecular clouds}

\citet{Jenkins09} found that atomic oxygen is depleted from the gas phase at a rate that cannot be explained by its presence in silicate and metallic oxides grains. 
Thus, some atomic oxygen must be sequestered into grains in a form other than silicate and metallic oxides; however, the corresponding reservoir is missing. The fraction of missing oxygen atoms with respect to the available oxygen atoms can reach $\sim28\%$ in translucent clouds \citep{Whittet10}. 

The missing oxygen might exist in the form of large water-ice grains, i.e., larger than $1~\mu$m \citep{Jenkins09, Poteet15}.
The presence of large ice grains does not contradict the absence of the $3.1$-$\mu$m absorption feature observed for translucent clouds, as large ice grains do not show a strong absorption feature. The presence of $\mu$m-sized ice grains is also consistent with the flat extinction curves observed at mid-infrared wavelengths \citep{Wang15}.
However, it has also been suggested that icy grains may have a short lifetime in harsh environments in interstellar media. Instead, organic matter may be an alternative reservoir of the missing oxygen \citep{Whittet10, Jones19}.

Observations of the polarization feature of translucent clouds may shed light on the reservoir of missing oxygen. Unlike the ice absorption feature, the polarization feature can arise for $\mu$m-sized icy particles, such as those shown in Figure \ref{fig:sizedist1}. Therefore, spectropolarimetric observations of scattered light from translucent clouds at the ice band will be useful for determining whether water ice is the reservoir.

\subsection{Molecular cloud cores}

Mid-infrared scattering has been observed for cloud cores, known as coreshine \citep{Steinacker10, Pagani10,Andersen13,Lefevre14,Steinacker14a,Steinacker14b,Steinacker15}, indicating the presence of $\mu$m-sized grains in cloud cores. Grain growth provides a potential explanation for coreshine, as not all cores show coreshine; the detection rate is approximately one half among 110 investigated cores \citep{Pagani10}. Molecular cloud cores may be sufficiently dense to initiate dust coagulation \citep{Ossenkopf93, Ormel09, Ormel11}; however, this process may require cores that are denser and more turbulent than current estimates \citep{Steinacker14b}. Thus, to reveal how grain growth proceeds in cores, it would be valuable to characterize icy grains in cores.

Observing the polarization feature will help to constrain both the grain-size distribution and ice abundance that causes coreshine because $\mu$m-sized ice-rich grains typically show a strong polarization feature (Figures \ref{fig:sizedist1} and \ref{fig:k99}). In addition, coreshine has been identified at both {\it Spitzer} wavelengths and in the K band \citep{Andersen13}. Thus, spectropolarimetry measurements of coreshine in the ice band may provide a promising target for the polarization feature.

\subsection{Protoplanetary disks}
\citet{Fukagawa10} reported that disk-scattered light at near-infrared wavelengths in protoplanetary disks is often gray, indicating the presence of aggregates with radii larger than a micron \citep{Mulders13, Tazaki19}. In addition, disk-scattered light primarily comes from the outer disk region, where water ice is expected to freeze out onto grains \citep[e.g.,][]{Chiang01}. Therefore, disk-scattered light is expected to show the scattering-polarization feature.

Although the 3-$\mu$m feature has been observed as an absorption feature for edge-on disks \citep{Pont05, Terada07, Honda09, Terada12a,Terada12b, Honda16, Terada17}, one can also detect the ice feature imprinted in disk-scattered light \citep{Inoue08}. In fact, a signature of the 3-$\mu$m feature in disk-scattered light has been detected via multi-color imaging \citep{Honda09, Honda16}. Because $\mu$m-sized icy grains exhibit a strong polarization feature, as shown in Figure \ref{fig:sizedist1}, protoplanetary disks represent another promising target for the polarization feature.

The water-ice observations in protoplanetary disks have been anticipated with the forthcoming space telescope, {\it JWST}, which will be capable of observing total intensity of scattered light of the 3-$\mu$m-ice feature. A more detailed study of both total and polarized intensity of the ice feature for protoplanetary disks will be presented in a forthcoming paper \citet{Tazaki_ice}.

\subsection{Debris disks} \label{sec:debris}

To our knowledge, a distinct water-ice feature has not yet been detected in debris disks \citep[see also][]{Hughes18}, although \citet{Chen08} reported a broad peak at $60-75~\mu$m that may be due to crystalline water ice.
Future space missions, such as {\it JWST}, and ground-based facilities, such as E-ELT and TMT, will advance our understanding of water ice in debris disks \citep{Kim19}.

In Section \ref{sec:amin}, we argued that the polarization feature may be a good indicator of $\amin$ for cases in which $q>3$ and $\amax\gg10~\mu$m. These conditions appear to be frequently satisfied in debris disks \citep[e.g.,][]{Hughes18}.
For example, based on {\it Spitzer} observations of silicate emission features observed for 120 debris disks, \citet{Mittal15} found that $\amin\sim0.3-40~\mu$m and $q=3.5-4.0$, which is a suitable parameter space for the polarization feature, as shown in Figure \ref{fig:sizedist1}. Therefore, we anticipate that spectropolarimetric observations of the ice feature will be useful for understanding icy grains in debris disks.

Recently, \citet{Kim19} reported that the scattering-polarization feature at $3~\mu$m is a useful tracer of ice abundance and of different ice destruction processes in debris disks. The results shown in Figure \ref{fig:sizedist2} are qualitatively consistent with those of \citet{Kim19}. In addition, the results obtained by \citet{Kim19} show characteristics of the polarization feature in the reflection regime. 
The discussion presented in Section \ref{sec:origin} may provide a physical explanation for these previous results (e.g., Figure \ref{fig:pwavel}).

\section{Summary} \label{sec:summary}

We have studied scattering polarization of the 3-$\mu$m water-ice feature, termed "the scattering-polarization feature." In this study, we have augmented the work by \citet{Pendleton90} to clarify the role of size distribution, scattering angle, and ice abundance. In particular, we have newly proposed that surface reflection plays an important role in the polarization feature of large icy grains. 
The polarization feature of large icy grains was shown to be consistent with polarimetric observations of the envelope of the protostar L1551 IRS 5.

The primary findings of this work are as follows.
\begin{enumerate}
\item The scattering-polarization feature is sensitive to the presence of $\mu$m-sized ice grains. The feature profile varies significantly between $\amax=0.3$ and $10~\mu$m (Figure \ref{fig:polwave}). For $\amax\lesssim10~\mu$m, the feature becomes more prominent for larger maximum grain radii. 

\item The polarization feature remains nearly constant for $\amax\gtrsim10~\mu$m when the slope of the size distribution $q>3$. When $q<3$ and $\amax\gtrsim10~\mu$m, the degree of polarization increases with $\amax$ for all ice band wavelengths (Figure \ref{fig:sizedist1}).

\item For $\amax\gtrsim10~\mu$m and $q>3$, the polarization feature is useful for inferring $\amin$. In this case, the feature is sensitive to $\amin$ values ranging from sub-micron to a few tens of microns (Figure \ref{fig:sizedist2}). As $\amin$ increases, the polarization degree increases for all ice band wavelengths.

\item Except for back-scattered light from ice-rich grains, the polarization feature is symmetric with respect to $\lambda\sim3~\mu$m for various ice abundances and scattering angles. In contrast, back-scattered light from ice-rich grains shows an asymmetric profile, and the peak polarization shifts below $3~\mu$m (Figure \ref{fig:polfice}).

\item For large icy grains, 
the polarization degree is enhanced at $\lambda\sim3\mu$m due to absorption inside the grains (Figures \ref{fig:regime} and \ref{fig:debyePOL}). The high optical thickness attenuates internal scattering and increases the contribution of surface reflection, which tends to be highly polarized due to Brewster scattering (Figure \ref{fig:pwavel}).

\item The polarization excess at $\lambda\sim3~\mu$m observed for the envelope of the low-mass protostar L1551 IRS 5 is consistent with scattering of $\mu$m-sized icy grains (Figure \ref{fig:k99}). Our model suggests the presence of grains with $\amax\gtrsim 3 \mu$m and a relatively low ice abundance ($\fice\lesssim 0.03$).

\end{enumerate}

With ground-based telescopes, it is difficult to access scattered light at wavelengths longer than $\sim3~\mu$m due to atmospheric emission as well as emission from instruments. Therefore, future space-based mid-infrared polarimetry will be important for elucidating the icy universe. 

\acknowledgments
R.T. acknowledges JSPS overseas research fellowship. This work was supported by JSPS KAKENHI grant numbers JP17H01103 (R.T., T.M., M.H) and JP19H05068 (R.T.).

\software{PlotDigitizerX: \url{http://www.surf.nuqe.nagoya-u.ac.jp/~nakahara/Software/PlotDigitizerX/index-e.html} }

\appendix

\section{Computation of the Debye series} \label{sec:debye}
\subsection{What is the Debye series?}
The Debye series is a geometric series expansion of scattering coefficients for a sphere with respect to reflection coefficients. Because the expansion is rigorous, the infinite series sum of the Debye series exactly recovers the Mie theory \citep{vdp37a, vdp37b, Hovenac92}.
This method bears the name of Peter Debye because he first introduced the geometric series expansion of scattering coefficients, although his first attempt was for an infinitely long circular cylinder under the short-wavelength approximation \citep{Debye1908}.

The concept of the Debye series is very similar to that of the geometrical optics approximation. Indeed, under some assumptions, the Debye series can be exactly reduced to the geometrical optics approximation \citep{vdh46, vdh57}. However, the Debye series has important benefits. For example, it can overcome many difficulties that arise in the geometrical optics approximation, such as the intensity divergence observed at rainbow and glory angles \citep{vdh57, Bohren83}, the negligence of surface waves \citep{Hovenac92}, and its limitation to sufficiently large spheres. Thus, we prefer to use the Debye series instead of the geometrical optics approximation. 

\subsection{Formulation}
The numerical calculation of the Debye series is much more complicated than that of the Mie theory. For this calculation, we adopt an algorithm developed by \citet{Shen10}. Here, we briefly summarize the procedures.
The scattering matrix elements $S_{11}$ and $S_{12}$ are defined by $S_{11}=(|S_1|^2+|S_2|^2)/2$ and $S_{12}=(|S_2|^2-|S_1|^2)/2$, respectively, where $S_1$ and $S_2$ are amplitude scattering matrix elements: 
\begin{eqnarray}
S_1&=&\sum_n \frac{2n+1}{n(n+1)}(a_n\pi_n+b_n\tau_n),\\
S_2&=&\sum_n \frac{2n+1}{n(n+1)}(a_n\tau_n+b_n\pi_n).
\end{eqnarray}
Here, $a_n$ and $b_n$ are scattering coefficients corresponding to TM and TE modes, respectively, and
$\pi_n=P_n^1/\sin\theta$ and $\tau_n=dP_n^1/d\theta$ are angle-dependent functions, with $P_n^1$ representing the associated Legendre function \citep{Bohren83}. The scattering coefficients are computed by the Debye series \citep{Hovenac92, Shen10}:
\begin{eqnarray}
\left.
\begin{array}{l}
a_n  \\
b_n 
\end{array} 
\right\}
&=&\frac{1}{2}\left[1-R_n^{212}-\frac{T_n^{21}T_n^{12}}{1-R_n^{121}}\right],\\
&=&\frac{1}{2}\left[1-R_n^{212}-\sum_{p=1}^{\infty} T_n^{21}(R_n^{121})^{p-1}T_n^{12}\right], \label{eq:basic1}
\end{eqnarray}
where $R_n^{212}$ and $R_n^{121}$ are partial wave reflection coefficients and $T_n^{12}$ and $T_n^{21}$ are partial wave transmission coefficients, defined as
\begin{eqnarray}
T_n^{21}&=&m\frac{2i}{\alpha\xi_n^{(1)'}(x)\xi_n^{(2)}(y)-\beta\xi_n^{(1)}(x)\xi_n^{(2)'}(y)},\label{eq:set1}\\
T_n^{12}&=&\frac{2i}{\alpha\xi_n^{(1)'}(x)\xi_n^{(2)}(y)-\beta\xi_n^{(1)}(x)\xi_n^{(2)'}(y)},\label{eq:set2}\\
R_n^{212}&=&\frac{\alpha\xi_n^{(2)'}(x)\xi_n^{(2)}(y)-\beta\xi_n^{(2)}(x)\xi_n^{(2)'}(y)}{\beta\xi_n^{(1)}(x)\xi_n^{(2)'}(y)-\alpha\xi_n^{(1)'}(x)\xi_n^{(2)}(y)},\label{eq:set3}\\
R_n^{121}&=&\frac{\alpha\xi_n^{(1)'}(x)\xi_n^{(1)}(y)-\beta\xi_n^{(1)}(x)\xi_n^{(1)'}(y)}{\beta\xi_n^{(1)}(x)\xi_n^{(2)'}(y)-\alpha\xi_n^{(1)'}(x)\xi_n^{(2)}(y)}. \label{eq:set4}
\end{eqnarray}
Here, $(\alpha,\beta)=(1,m)$ and $(m,1)$ for TE and TM waves, respectively; $x$ is a size parameter; $y=mx$; $\xi_n^{(1)}(z)=zh_n^{(1)}(z)$ and $\xi_n^{(2)}(z)=zh_n^{(2)}(z)$ are Riccati-Bessel functions; and $h_n^{(1)}$ and $h_n^{(2)}$ are spherical Hankel functions. We use the algorithm developed by \citet{Shen10} to solve Equations (\ref{eq:set1}) - (\ref{eq:set4}).

The first term of Equation (\ref{eq:basic1}) is $a_n=b_n=1/2$, which corresponds to Fraunhofer diffraction \citep{vdh46, vdh57}. The second term corresponds to external reflection, and the summation in the third term represents the internal reflection. 
It is convenient to define scattering coefficients for each component. For surface scattering ($p=0$ and diffraction), we have
\begin{eqnarray}
\left.
\begin{array}{l}
a_n^{(\mathrm{surf})}  \\
b_n^{(\mathrm{surf})}   
\end{array} 
\right\}=\frac{1}{2}(1-R_n^{212}), \label{eq:surf}
\end{eqnarray}
and for $p\ge1$, we have
\begin{eqnarray}
\left.
\begin{array}{l}
a_n^{(p)}  \\
b_n^{(p)}   
\end{array} 
\right\}=-\frac{1}{2}T_n^{21}(R_n^{121})^{p-1}T_n^{12}. \label{eq:comp}
\end{eqnarray}
By using $a_n^{(p)},~b_n^{(p)}$ (with $p=\mathrm{surf}, 1, 2, \dots$), we can define $S_1^{(p)}$ and $S_2^{(p)}$ and subsequently $S_{11}^{(p)}$ and $S_{12}^{(p)}$. For the surface-scattering component, we combine Fraunhofer diffraction and externally reflected light because this approach is more numerically stable than the case in which these terms are treated separately.

\subsection{Benchmark test}
\begin{figure*}[t]
\begin{center}
\includegraphics[width=0.49\linewidth,keepaspectratio]{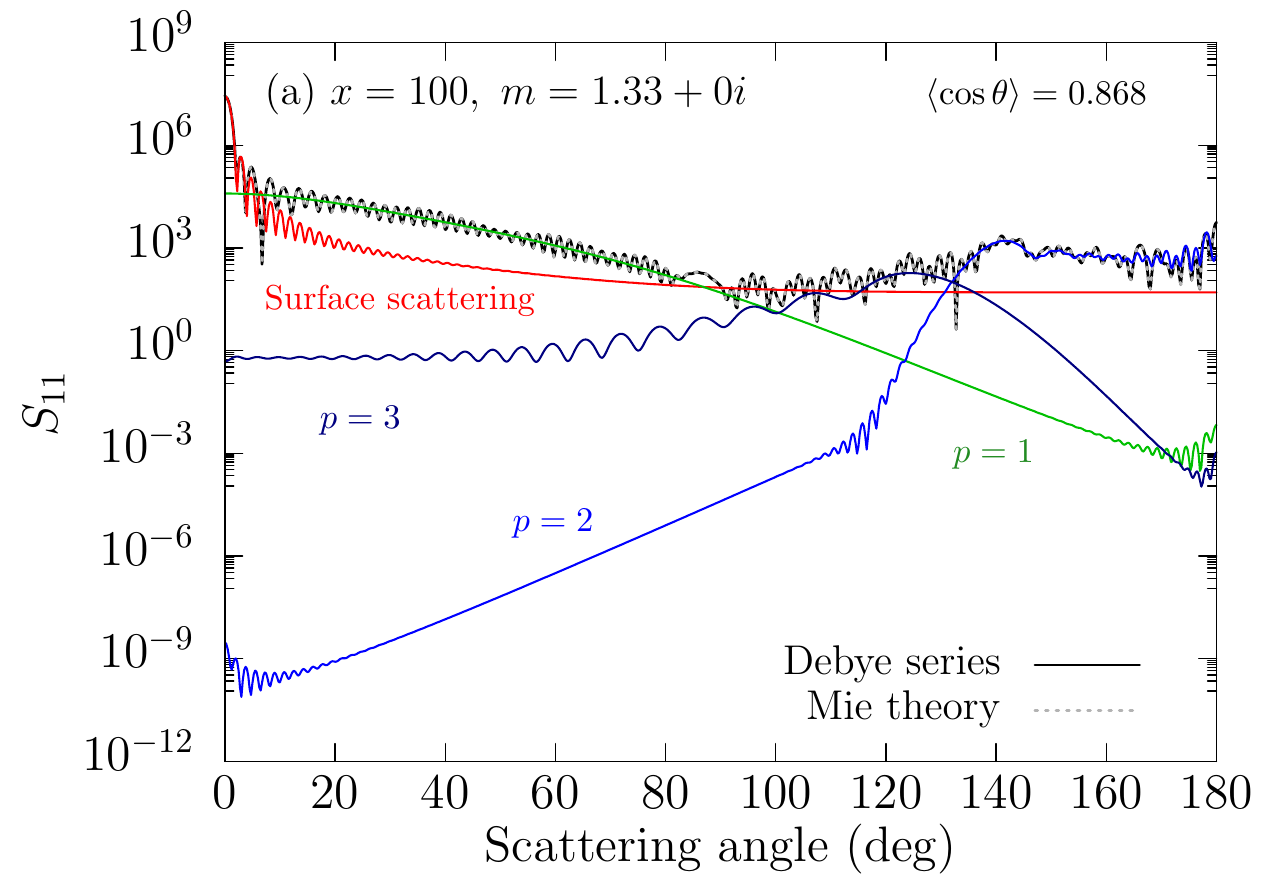}
\includegraphics[width=0.49\linewidth,keepaspectratio]{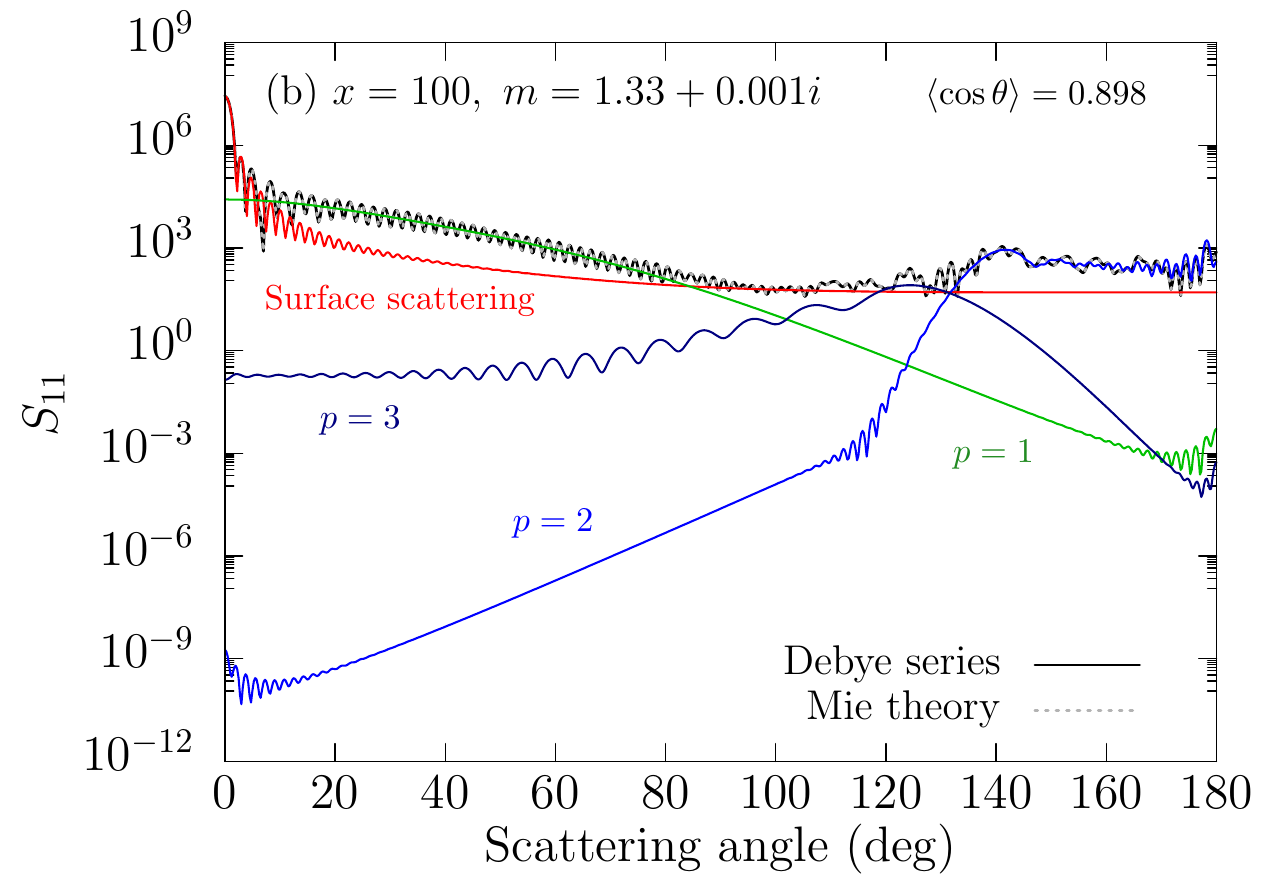}
\includegraphics[width=0.49\linewidth,keepaspectratio]{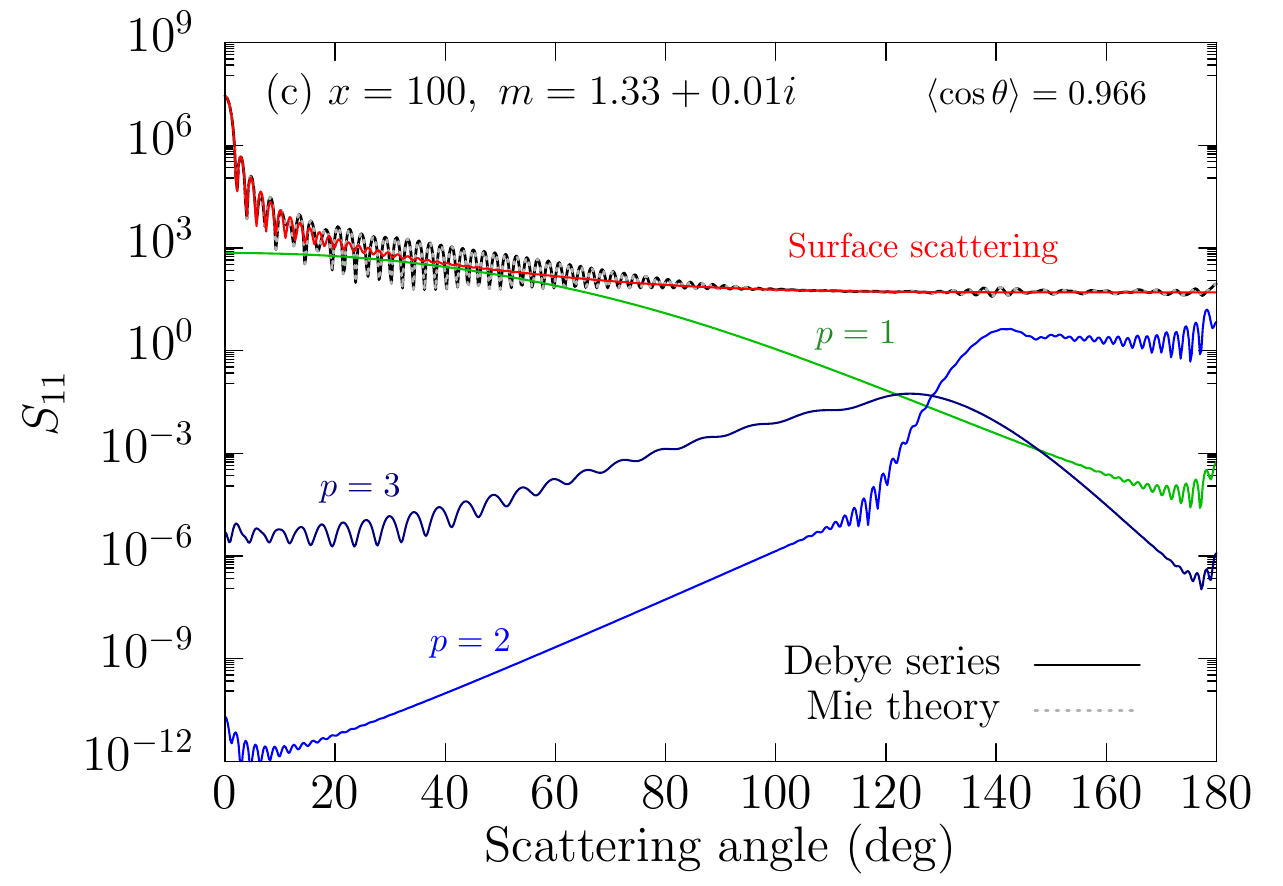}
\includegraphics[width=0.49\linewidth,keepaspectratio]{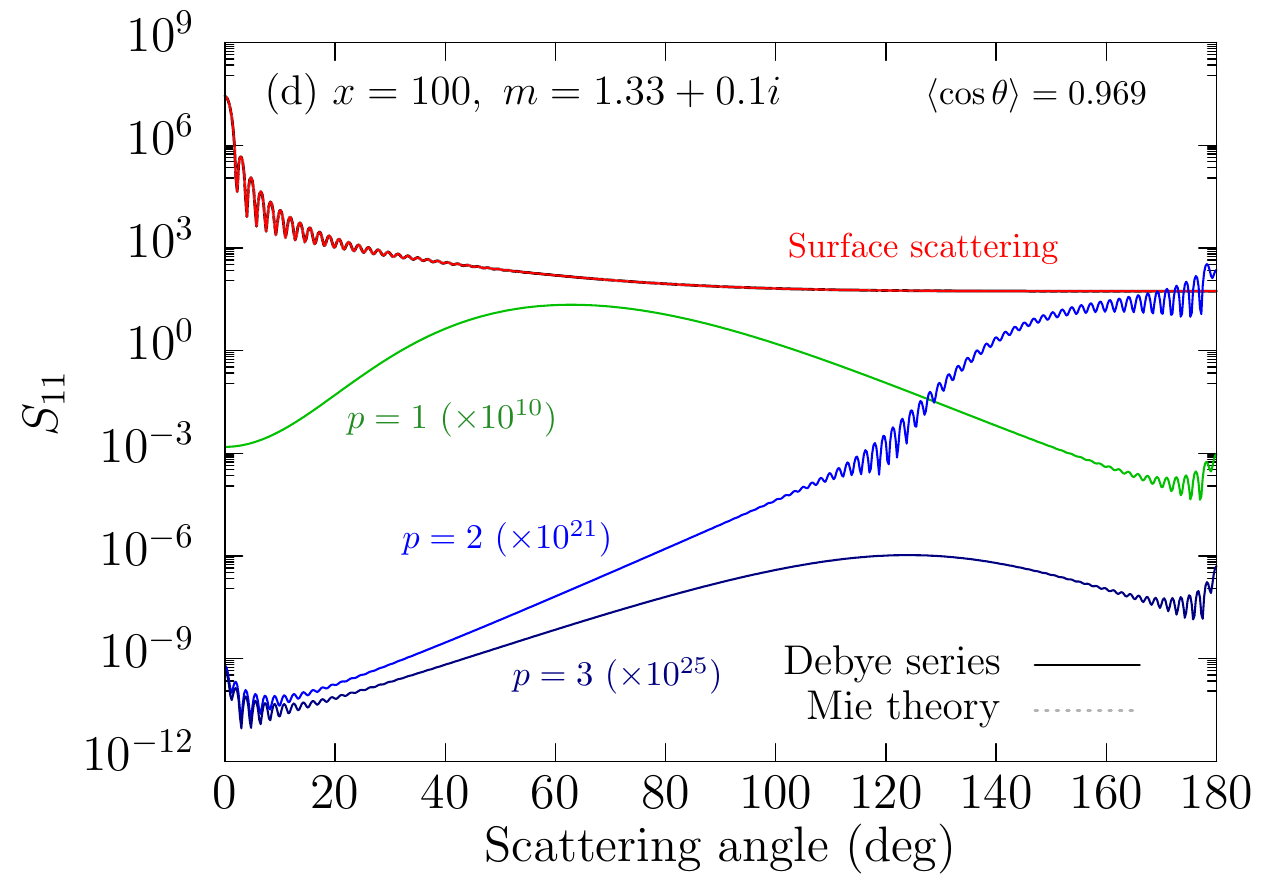}
\caption{Benchmark calculations of the Debye series. The black solid and gray dashed lines show results obtained by the infinite sum of the Debye series and the Mie theory, respectively. Each panel shows results for a different refractive index: (a) $m=1.33+0i$, (b) $m=1.33+0.001i$, (c) $m=1.33+0.01i$, and (d) $m=1.33+0.1i$.
The solid lines show each component of the Debye series: surface-scattered light $S_{11}^{(\mathrm{surf})}$ (red), twice-refracted light $S_{11}^{(p=1)}$ (green), light subjected to one internal reflection $S_{11}^{(p=2)}$ (light blue), and light subjected to two internal reflections $S_{11}^{(p=3)}$ (dark blue). For panel (d), the components with $p\ge1$ are multiplied by the number indicated in the panel.}
\label{fig:bench}
\end{center}
\end{figure*}

\begin{figure*}[t]
\begin{center}
\includegraphics[width=0.49\linewidth,keepaspectratio]{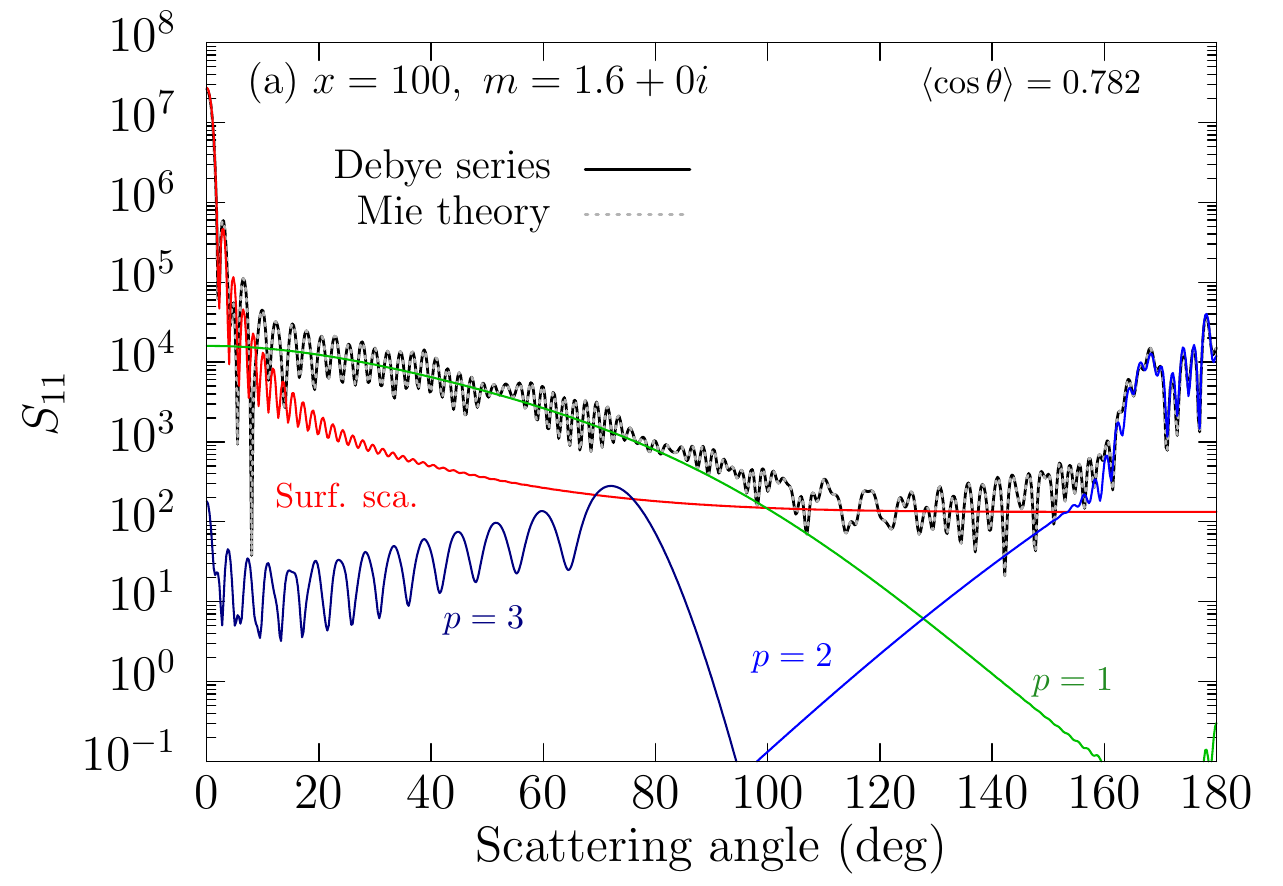}
\includegraphics[width=0.49\linewidth,keepaspectratio]{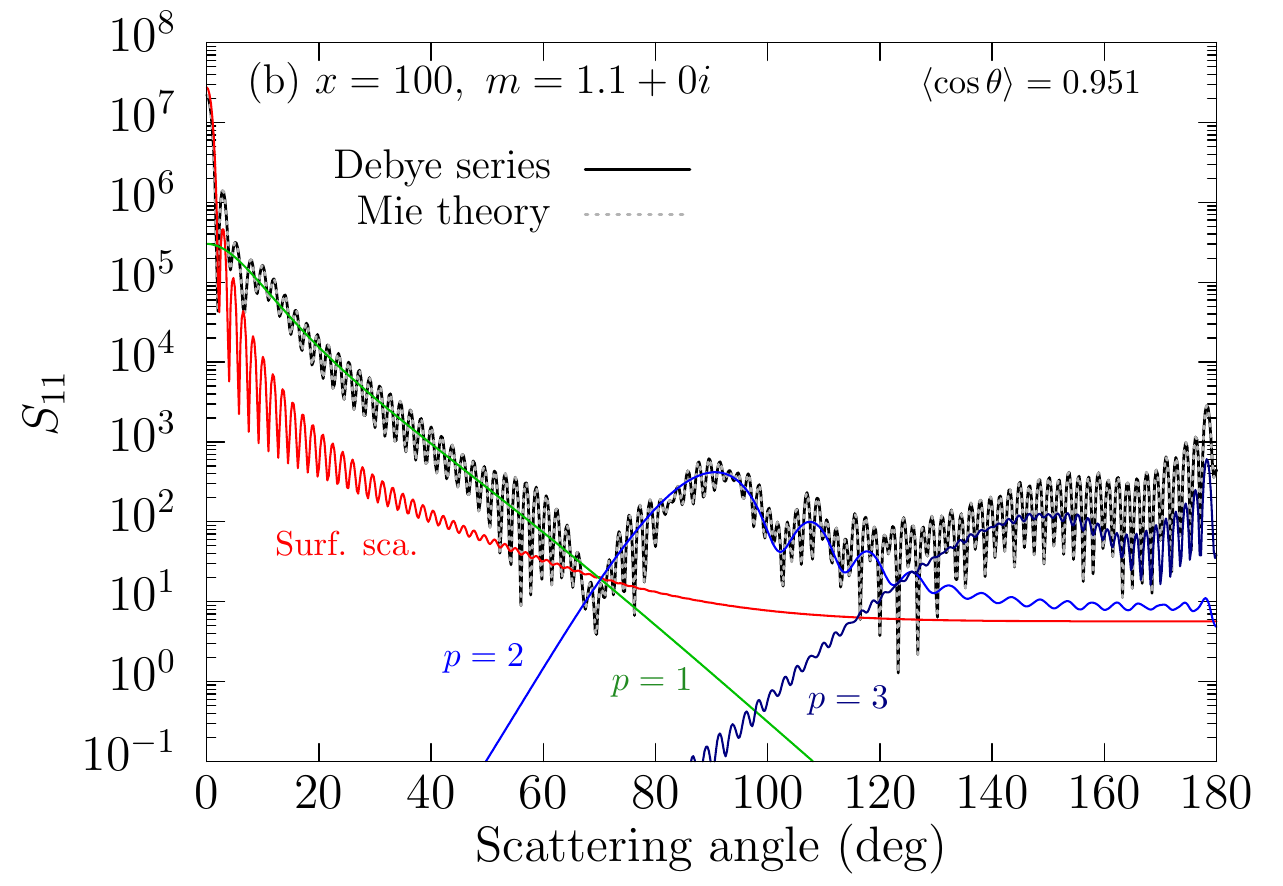}
\caption{Same as Figure \ref{fig:bench}, but for different values of the real part of the refractive index. (a) $m=1.6+0i$. (b) $m=1.1+0i$.}
\label{fig:bench2}
\end{center}
\end{figure*}

First, we confirm that the scattering properties obtained by summing the Debye series to infinity can successfully recover those obtained by the Mie theory (see Figures \ref{fig:debyeS11}, \ref{fig:bench}, and \ref{fig:bench2}). 
Next, we compute the optical properties of single-sized spheres with size parameter $x=100$ and refractive index $m=n+ik$, where $n=1.33$ and $k=0$, $0.001$, $0.01$, and $0.1$. The results are shown in Figure \ref{fig:bench}. These results were compared with those presented in \citet{Shen10} and showed good agreement. 

Each figure shows the asymmetry parameter of the phase function, $\langle \cos\theta \rangle$. This parameter increases as the imaginary part of the refractive index increases because higher-order scattering ($p\ge1$) tends to be suppressed for highly absorbing grains. The asymmetry parameter increases as the real part of the refractive index decreases (Figure \ref{fig:bench2}) because scattered light with $p=1$ is likely confined to small scattering angles.

\bibliography{cite}

\end{document}